\documentclass[acmsmall, nonacm]{acmart}

 \AtBeginDocument{%
  }

\usepackage{graphicx, amsmath, mathtools, proof, listings} 
\usepackage{float,caption,subcaption,booktabs,multirow, multicol,array}
\usepackage{dutchcal}
\usepackage{hyperref, csquotes}
\usepackage[noabbrev,capitalise]{cleveref}
\usepackage{tikz, tikz-cd}
\usetikzlibrary{arrows,decorations.pathmorphing,shapes.misc}

\lstnewenvironment{code}
{\lstset{
basicstyle=\footnotesize\ttfamily,
mathescape=true,
escapeinside={(*@}{@*)},
tabsize=2,
numbers=left,               
stepnumber=1,    
breaklines=true,
xleftmargin=5.0ex,
numberfirstline=true,
literate={\ \ }{{\ }}1}}{}

\usepackage{adjustbox}

\usepackage{silence}
\newcommand{\name}{OptiFusion}
\newcommand{\syntax}[1]{\mathtt{#1}}

\newcommand*{\AGG}             {\ensuremath{\mathbcal{S}_{op}^{s}}}
\newcommand*{\RAGG}             {\ensuremath{^{remote}\mathbcal{S}_{op}^{s}}}
\newcommand*{\CAGG}             {\ensuremath{^{cached}\mathbcal{S}_{op}^{s}}}
\newcommand*{\LAGG}             {\ensuremath{^{local}\mathbcal{S}_{op}^{s}}}

\newcommand{\nop}[1]{}

\newcommand{\changem}[1]{\color{black}{#1}\color{black}}
\newcommand{\changea}[1]{\color{black}{#1}\color{black}}

\newcommand{\changec}[1]{\color{black}{#1}\color{black}}
\newcommand{\changetxtm}[1]{\textcolor{black}{#1}}
\newcommand{\changetxta}[1]{\textcolor{black}{#1}}
\newcommand{\changetxtb}[1]{\textcolor{black}{#1}}
\newcommand{\changetxtc}[1]{\textcolor{black}{#1}}

\setcopyright{none}
\renewcommand\footnotetextcopyrightpermission[1]{} 

\title[OptiFusion]{Using Process Calculus for Optimizing Data and Computation Sharing in Complex Stateful Parallel Computations}
\author{Zilu Tian}
\affiliation{
    \institution{University of Zurich}
    \city{Zurich}
    \country{Switzerland}
}
\email{zilu.tian@uzh.ch}
\orcid{0000-0003-4339-1876}

\author{Dan Olteanu}
\affiliation{
    \institution{University of Zurich}
    \city{Zurich}
    \country{Switzerland}
}
\email{dan.olteanu@uzh.ch}
\orcid{0000-0002-4682-7068}

\author{Christoph Koch}
\affiliation{
    \institution{EPFL}
    \city{Lausanne}
    \country{Switzerland}
}
\email{christoph.koch@epfl.ch}
\orcid{0000-0002-9130-7205}

\begin{document}
\begin{abstract}
We propose novel techniques that exploit data and computation sharing to improve the performance of complex stateful parallel computations, like agent-based simulations. Parallel computations are translated into behavioral equations, a novel formalism layered on top of the foundational process calculus $\pi$-calculus. Behavioral equations blend code and data, allowing a system to easily compose and transform parallel programs into specialized programs. We show how optimizations like merging programs, synthesizing efficient message data structures, eliminating local messaging, rewriting communication instructions into local computations, and \changetxtc{aggregation pushdown} can be expressed as transformations of behavioral equations. We have also built a system called OptiFusion that implements behavioral equations and the aforementioned optimizations. Our experiments showed that OptiFusion is over 10$\times$ faster than state-of-the-art stateful systems benchmarked via complex stateful workloads. Generating specialized instructions that are impractical to write by hand allows OptiFusion to outperform even the hand-optimized implementations by up to 2$\times$.
\end{abstract}

\maketitle 

\section{Introduction}
\label{sec:intro}
{Complex stateful parallel computations} refer to heterogeneous parallel computations on the bulk-synchronous parallel (BSP) machine~\cite{valiant90BSP}, where each computational unit can execute distinct code and perform in-place updates to its state~\cite{tian23generalizing}. The term \emph{complex} emphasizes the distinction with homogeneous stateful parallel computations, like in the vertex-centric paradigm, where all computational units execute the same code. Despite its simplicity, the vertex-centric paradigm is used by many popular distributed graph analytical systems, including Pregel~\cite{pregel}, Giraph~\cite{facebook-giraph}, GraphX~\cite{graphx14}, and Flink Gelly~\cite{flink-pregel}. In these systems, users specify the behavior of each vertex in an input graph by means of code. Each vertex corresponds to a computational unit. All vertices execute the same code. 

\nop{The computational model is based on the BSP model~\cite{valiant90BSP}. Parallel computational units proceed in a sequence of supersteps, separated by global synchronizations. Computational units interact by sending messages. Per superstep, each unit independently processes messages received from other units, updates its state, and sends messages to others. Messages are collected and delivered at the end of a superstep.}

Agent-based simulations are a prime example of complex stateful parallel computations. These simulations consist of concurrent agents interacting within a virtual world, each with its own state and code. They are highly flexible and have been extensively adopted across diverse fields in the social sciences, such as economics and epidemiology, where they serve as crucial tools for modeling complex phenomena~\cite{uk-covid, simulations-covid, econ-abm, econ2-abm, tian23generalizing}.

The computational model of agent-based simulations is based on the BSP model. Agent computations proceed in a sequence of rounds, separated by global synchronizations. Agents interact by sending messages. Per round, each agent independently processes messages received from other agents, updates its local state, and sends messages to other agents. Messages are collected and delivered by the underlying agent-based simulation framework at the end of a round and arrive at the mailbox of the receiving agents at the beginning of the following round.

\begin{figure}
    \centering
    \includegraphics[width=\columnwidth]{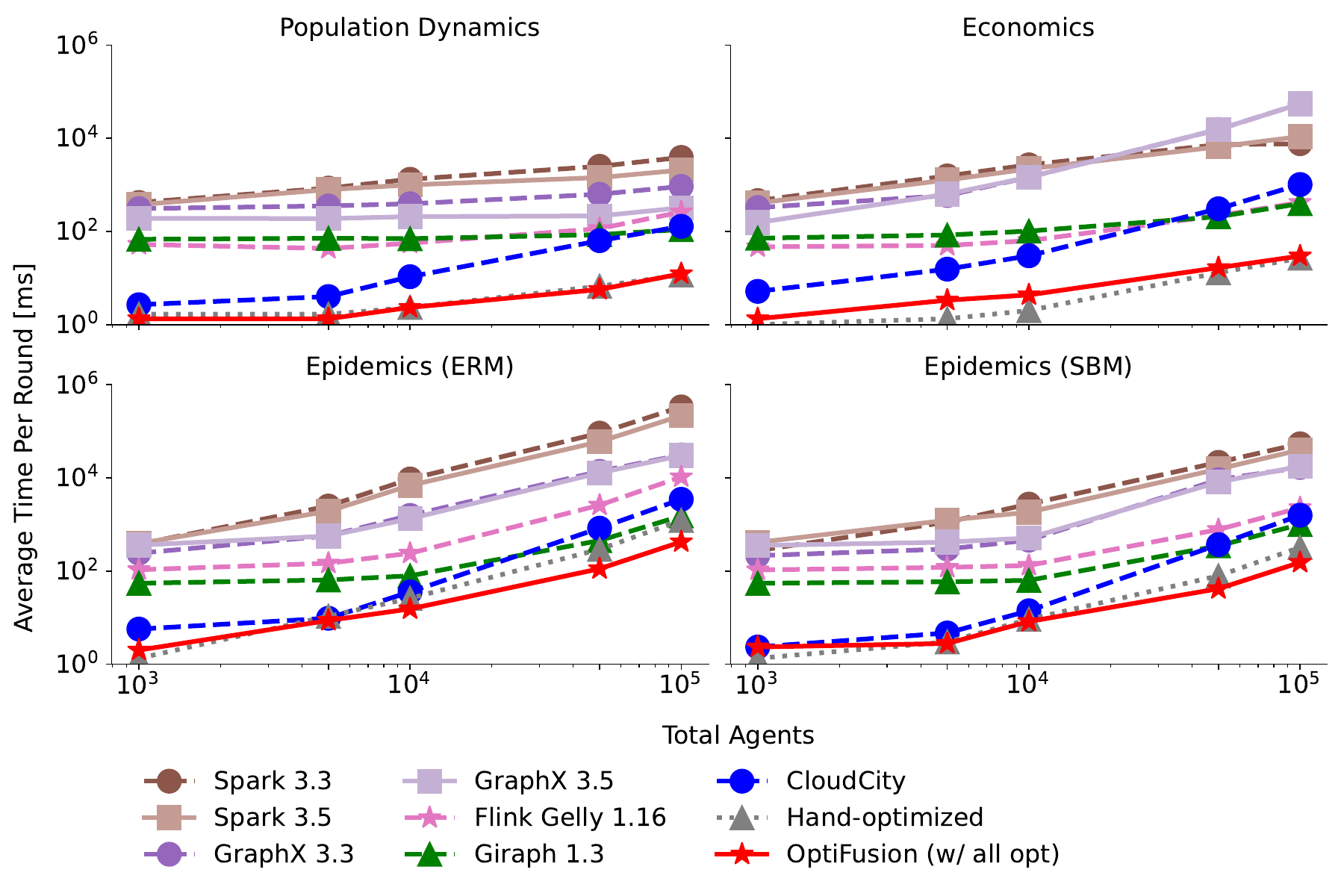}
    \caption{
    The performance gap between stateless parallel systems (Spark and GraphX) and stateful parallel systems (Flink, Giraph, and CloudCity) when benchmarked using complex stateful workloads remains. The workloads are simulations for population dynamics, economics, and epidemics where the social graph is generated from the Erd\H{o}s-R\'enyi random graph model (abbreviated ERM) and the stochastic block random model (abbreviated SBM) respectively. These experiments reproduce the scale-up results in \cite{tian23generalizing} using the latest software versions. By exploiting data and computation sharing optimizations, {\name} is over 500$\times$ faster than stateless systems and 10$\times$ faster than stateful systems. The performance of {\name} is comparable with and can even be better than the hand-optimized implementations by 2$\times$.}
    \label{fig:cross-sys}
\end{figure}

Prior work~\cite{tian23generalizing} showed that stateful parallel systems can outperform stateless parallel systems, such as Spark~\cite{spark} and GraphX~\cite{graphx14}, by up to 100$\times$ when benchmarked on agent-based simulations, \changetxta{owing to different system design features like in-place updates.} At the time of writing, Spark and GraphX have been updated from version 3.3 to 3.5.\footnote{Giraph has retired~\cite{apache-giraph}, and Gelly only supports Flink up to version 1.16~\cite{link-flink-gelly}.} 
A natural question is whether the latest version of Spark and GraphX still suffer from a similar performance degradation compared with stateful systems when benchmarked using the same complex stateful workloads, which leads to the experiments presented in \cref{fig:cross-sys}.

In \cref{fig:cross-sys}, we repeat the scale-up experiments in \cite{tian23generalizing} using the latest software and show the results for Spark (versions 3.3 and 3.5),
GraphX (versions 3.3 and 3.5),
Flink Gelly (version 1.16),
Giraph (version 1.3), 
and CloudCity (SNAPSHOT-2.0).\footnote{The slight performance variation compared with \cite{tian23generalizing} is due to different hardware configurations: the base clock frequency of our server is 2.2 GHz, but 2.7 GHz in \cite{tian23generalizing}.} As the number of agents increases, the previously reported performance gap between stateless and stateful parallel systems remains evident across all workloads in the benchmark. 

\Cref{fig:cross-sys} also contains experimental results of our system {\name} evaluated against other systems and the hand-optimized implementations. {\name} can be 10x faster than state-of-the-art stateful parallel systems like Giraph, Flink Gelly, and CloudCity. Furthermore, {\name} generates special instructions for agents in different graph components, which are impractical to write by hand as the number of agents within a component increases, by applying partial evaluation. This allows {\name} to outperform even hand-optimized implementations by up to 2$\times$ as the number of agents increases.

\changetxta{{\name} is a program specialization framework that compiles a generic agent program that is easy to define, into specialized programs that are efficient to execute, through multiple phases of program rewriting}. \changetxtb{Such rewriting is based on a novel intermediate representation for parallel programs called \emph{behavioral equations}, grounded in a foundational process calculus, $\pi$-calculus~\cite{milner89Communication, milner1993polyadic,milner99pi}. In formulating behavioral equations, we prioritize objectives listed in the desiderata below.}
\begin{itemize}

    \item{\emph{Expressing Non-deterministic Parallel Behavior}.}
    The language should be able to express system optimizations that exploit the non-deterministic execution order of parallel agents, such as merging agents to reduce the degree of parallelism to match available hardware parallelism in a system for better performance.
    
    \item \emph{Expressing Agent Computations}.
    The language should have high-level primitives that allow users to express the computation of individual agents declaratively. This simplifies optimizations that compose, reorder, and distribute computations, such as transforming computations by exploiting algebraic properties like associativity and commutativity.
    
    \item \emph{Expressing Agent Interactions}.
    Fine-grained interactions are central to the semantics of parallel agents. This requirement allows for optimizations that transform the communication pattern of agents, like transforming computations for processing messages into local computations over synthesized message data structures.

    \item \emph{Expressing Data and Computation Placement}. 
    The language should be able to express data and computation placement in a distributed environment, enabling locality-based optimizations such as \changetxtc{aggregation pushdown}, where data is combined locally on each computer in a distributed system and partial results are aggregated.
\end{itemize}

To summarize, this paper makes the following contributions.
\begin{itemize}
    \item We introduce a formalism named \emph{behavioral equations} based on $\pi$-calculus for expressing and optimizing complex stateful parallel computations. We also introduce annotations to express data and computation placement. 
    The syntax and semantics of the language are presented in \cref{sec:pm}. 
    \item We demonstrate the \changetxta{generality} and usability of behavioral equations by expressing a host of data-sharing and computation-sharing optimizations as transformations of behavioral equations in \cref{sec:optimizations}. 
    \item We build {\name}, a compile-time program specialization framework based on behavioral equations. The optimizer in {\name} exploits the aforementioned optimizations to generate specialized programs. We explain the system design and implementation details in \cref{sec:implementation}.
    \item Finally, we evaluate the effectiveness of program specializations in \cref{sec:eval}. Our experiments show that the performance of the generated programs is on-par with or up to 2$\times$ faster than hand-optimized programs for all workloads in an agent-based simulation benchmark. {\name} can be 50$\times$ faster than other systems like CloudCity, \changetxtc{Giraph, and Flink Gelly}.
\end{itemize}
\section{Behavioral Equations}
\label{sec:pm}
This section introduces \emph{behavioral equations}, a language based on the $\pi$-calculus for modeling complex stateful parallel computations in bulk-synchronous parallel (BSP) systems~\cite{valiant90BSP}. To ease the formulation of behavioral equations, we start by explaining the BSP model and \changetxtm{presenting a gentle introduction to $\pi$-calculus}, before detailing the syntax and semantics of behavioral equations. We also introduce partition-annotated behavioral equations, which express data and computation placement by annotations. 

\subsection{BSP Model}
The BSP model is an abstract parallel machine that underpins many distributed frameworks designed for efficient and scalable parallel computing~\cite{spark, graphx14, facebook-giraph, pregel, apache-flink, tian23generalizing}. This abstract machine consists of:
\begin{itemize}
    \item A set of processors. Each processor is a core-memory pair with private memory. A processor updates values stored in the memory locally. In addition, processors can communicate via sending and receiving messages.
    \item A synchronization facility to synchronize all processors periodically. 
    Synchronizations divide the parallel computation of processors into a sequence of \emph{supersteps}. Per superstep, a processor performs arbitrarily many \emph{steps}, in the form of updating local values, processing received messages, and sending messages. Messages are delivered at the end of a superstep and arrive at the beginning of a superstep.
\end{itemize}
For simplicity, we assume messages arrive at the beginning of the next superstep in our discussion.

\subsection{Pi Calculus Primer}
\label{subsec:pi-calculus}
The \emph{$\pi$-calculus} provides a mathematical framework for modeling interactions in concurrent systems. There are two basic entities, {names} and {processes}.\footnote{Many versions of $\pi$-calculus exist~\cite{milner1993polyadic, Milner92Mobile1, milner89Communication}. This introduction is based on~\cite{milner1993polyadic}.} A \emph{name} represents a communication channel (abbreviated as channel) or value. A \emph{process} interacts with other processes by sending and receiving names along channels. Processes can be composed in parallel with other processes, create new names, and spawn new processes. \changetxtm{Below, we present the syntax and semantics of $\pi$-calculus, followed by a concrete example.}

\subsubsection{Syntax}
Let $X$ be the set of names, $x, a \in X$. A {process} is built from names, using \emph{prefix actions} of the form $a(x)$ and $\overline{a}x$, and operators for \emph{choice}, \emph{parallel composition}, \emph{restriction}, and \emph{replication}. These operators are represented by symbols $+$, $|$, $\nu$, and $!$, respectively. More concretely, the syntax of process expressions $P$ and $Q$ is defined by the following BNF grammar rule:
$$
P, Q ::= 0 \mid  \overline{a}x.P \mid a(x).P \mid P + Q \mid P | Q \mid (\nu a)P  \mid\ !P  
$$
where
\begin{itemize}
    \item $0$ represents the inactive process.
    \item $\overline{a}x.P$ is a process waiting to send name $x$ on the output channel $\overline{a}$ (bar represents the outgoing direction) before continuing as process $P$.
    \item $a(x).P$ is a process waiting to receive a name on the input channel $a$ before binding the received name to $x$ in $P$ and proceeding as process $P$. If the name $x$ is not free in $P$, then alpha-conversion is required. 
    \item $P + Q$ is a process that can take part in either $P$ or $Q$ for communication. The process does not commit to any alternative until communication happens; once it does, the occurrence precludes the other alternative.
    \item $P | Q$ is a process that consists of $P$ and $Q$ that run in parallel.
    \item $!P$ replicates arbitrarily many copies of process $P$.
    \item $(\nu a)P$ creates a new name $a$ that is known only by the prefixed process $P$.
\end{itemize}

\changem{
\subsubsection{Equivalence}
\begin{figure}
    \centering
\begin{align*}
\textbf{(CHOICE-COMM)} & \quad P + Q \equiv Q + P \\
\textbf{(CHOICE-ASSOC)} & \quad (P + Q) + R \equiv P + (Q + R) \\
\textbf{(CHOICE-IDENT)} & \quad P + 0 \equiv P \\
\textbf{(PAR-COMM)} & \quad P \mid Q \equiv Q \mid P \\
\textbf{(PAR-ASSOC)} & \quad (P \mid Q) \mid R \equiv P \mid (Q \mid R) \\
\textbf{(PAR-IDENT)} & \quad P \mid 0 \equiv P \\
\textbf{(RES-SWAP)} & \quad (\nu a)(\nu b)P \equiv (\nu b)(\nu a)P \\
\textbf{(RES-SCOPE)} & \quad (\nu a)(P \mid Q) \equiv P \mid (\nu a)Q \quad \text{if } a \notin \text{fn}(P) \\
\textbf{(RES-ANN)} & \quad (\nu a)0 \equiv 0 \\
\textbf{(REPLICATION)} & \quad !P \equiv P \mid !P \\
\textbf{(ALPHA-CONV)} & \quad P \equiv P[a/x] \quad \text{if } a\notin \text{name}(P), x \in \text{bn}(P).
\end{align*}
    \caption{Structural congruence rules for $\pi$-calculus.}
    \label{fig:pi-cong}
\end{figure}
To simplify the presentation of semantics, we introduce an equivalence relation (denoted $\equiv$) between processes --  structural congruence~\cite{milner1993polyadic} -- defined by the rules in \cref{fig:pi-cong}.

Observe that Rules \textbf{RES-SCOPE} and \textbf{ALPHA-CONV} contain references to $fn(P)$, $bn(P)$, and $name(P)$. The first two are functions that return the set of names in process $P$ that are \emph{free} and \emph{bound}  respectively. A name $x$ is bound if it appears in the restriction operator $\nu x$ or as the name for the received value of an input channel $a(x)$ in the prefix action; otherwise, it is free. 
\nop{For prefix actions, we have: 
$$
fn(a(x)) = \{a \}, \ bn(a(x)) = \{ x\}, \ fn(\overline{a}x) = \{a, x\}, \ bn(\overline{a}x) = \{ \}.
$$}
The function $name(P)$ returns the set of all names, $name(P) = fn(P) \cup bn(P)$.

We briefly explain the rules in \cref{fig:pi-cong}. Operators $+$ and $|$ are commutative and associative (\textbf{CHOICE-COMM}, \textbf{CHOICE-ASSOC}, \textbf{PAR-COMM}, \textbf{PAR-ASSOC}). Interacting with the inactive process 0 does not change the behavior of the other process (\textbf{CHOICE-IDENT}, \textbf{PAR-IDENT}). New names generated for the same prefixed process can be swapped (\textbf{RES-SWAP}). The scope of a new name for two parallel processes can be restricted to one process if the name is bound in the other process (\textbf{RES-SCOPE}). Creating a new name in the inactive process has no effect (\textbf{RES-ANN}). 
The replication operator can create a new process (\textbf{REPLICATION}). Finally, two processes are structurally congruent if they only differ by a change of bound names (\textbf{ALPHA-CONV}).
$P[a/x]$ denotes the process $P$ in which name $a$ is substituted for the bound name $x$.

\subsubsection{Semantics}
We explain the meaning of $\pi$-calculus process expressions with reduction rules. We use $P \rightarrow Q$ to denote that \emph{process $P$ can perform a {computation step} and be transformed into process $Q$}. Every computation step in $\pi$-calculus requires the interaction between two processes, captured by the communication rule~\cite{milner1993polyadic}:
$$
\textbf{(Communication)} \quad (\ldots + \overline{a}x.P) \mid (\ldots + a(y).Q) \to P \mid Q[x/y]
$$
The communication rule has two key aspects. First, communication occurs between two {complementary} parallel processes, where one is waiting to send a name $x$ along the output channel $\overline{a}$ and the other is waiting to receive a name along the input channel with the same name $a$. Second, once communication occurs, other possible communications -- shown as $\ldots$ in the rule -- are discarded.

The communication rule is the only axiom; other reduction rules are inference rules: 
\[
\begin{array}{ccc}
\infer
    {P \mid Q \to P' \mid Q}
    {P \to P'}
& 
\infer
    {(\nu x)P \to (\nu x)P'}
    {P \to P'}
& 
\infer
    {Q \to Q'}
    {Q \equiv P \quad P \to P' \quad P' \equiv Q'}
\end{array}
\]
From left to right, the rules state that reductions can occur under (a) parallel composition and (b) restriction. Additionally, (c) structurally congruent processes have the same reduction.

\subsubsection{Example}
For a minimal example, we model the abstract behavior of a DRAM cell, the basic unit of computer memory: a DRAM cell can store one value; reading is destructive and erases the cell's current value. The cell behavior can be modelled as a process $B$: 
\begin{equation*}
B \stackrel{def}= i(x).(\overline{o}x.B + B)
\end{equation*}
By the convention of $\pi$-calculus, RHS is a process expression and LHS is a process identifier. A process identifier is purely syntactic and is substituted by the RHS expression during reductions.

Process $B$ waits to receive a name along the input channel $i$ and substitutes the received name for the bound name $x$ in the prefixed expression. Once it receives the value, $B$ has two options: (a) sends the received name along the output channel $\overline{o}$ before continuing as $B$, and (b) continues as $B$.

We now demonstrate how interactions transform $\pi$-calculus processes through a sequence of reductions. Assume a user process that writes 5 and 6 sequentially to an empty cell and reads from it. A system that consists only of the user process and an empty cell is modelled as follows. The $\nu$ operator ensures that names $i$ and $o$ are known only to the user process and the cell. Reduction arrows are labelled with interacting actions for clarity.
\begin{align*}
& (\nu i)(\nu o)(i(x).(\overline{o}x.B + B) \mid \overline{i}5.\overline{i}6.o(x).0) \\
\equiv &\ i(x).(\overline{o}x.B + B) \mid \overline{i}5.\overline{i}6.o(x).0 \\
\xrightarrow{i(x) \mid \overline{i}5} & \ (\overline{o}5.B + i(x).(\overline{o}x.B + B)) \mid \overline{i}6.o(x).0 \\
\xrightarrow{i(x) \mid \overline{i}6} & \ (\overline{o}6.B + B) \mid o(x).0 \\
\xrightarrow{\overline{o}6 \mid o(x)} & \ B \mid 0 \equiv B
\end{align*}

The example demonstrates how to express concurrent behavior and fine-grained interactions using $\pi$-calculus. However, expressing functions that describe high-level computations is nontrivial and often results in convoluted expressions~\cite{milner90functions}. We address this by behavioral equations, which lets the users program using high-level abstractions like \emph{states} and \emph{functions}. A behavioral equation is a {macro} that generates $\pi$-calculus processes.
}

\subsection{Syntax of Behavioral Equations}
A \changetxta{\emph{state}} $p$ models the core-memory pair in a BSP processor. It consists of a value and a unique state name. The basic building block for programming {states} is called \emph{behavioral equation}.

Intuitively, for a state $p$, users should be able to specify how to transit to another state declaratively, based on values of $p$ and optionally other states. We capture this programming model using \emph{behavior equations}, expressed as:
\changea{
\begin{equation}
p \xrightarrow{f\{i_1, \ldots, i_n\}} q,
\label{eq:behEqTrans}
\end{equation}
}
where $p, q$ are state names, $f$ is a user-defined function for computing the value of $q$, and $\{i_1, \ldots, i_n\}$ is a reference set that contains state names $i_1$ to $i_n$ needed for computing $q$. Equivalently, we can express \cref{eq:behEqTrans} in the alternative representation below:
\begin{equation}
p:=f\{i_1, \ldots, i_n\}.q,
\label{eq:beh}
\end{equation}
following the $\pi$-calculus convention. The symbol $:=$ means \emph{is defined as}. Unlike a {process identifier}, a state name on the LHS cannot be substituted by the expression on the RHS during reductions. The equation reads \enquote{\emph{The behavior of state $p$ is to obtain values from states $i_1, \ldots, i_n$, apply function $f$ over the obtained values and its value, initialize state $q$ with the computed result, and behave like state $q$.}}
The interactions among states are made precise using $\pi$-calculus. Behavioral equations let users program with high-level abstractions like \emph{functions} and \emph{states} in $\pi$-calculus.
\nop{as summarized below.
\begin{center}
    \resizebox{0.65\columnwidth}{!}{\begin{tikzpicture}[x=0.75pt,y=0.75pt,yscale=-1,xscale=1]

\draw   (184.06,64) -- (317,64) -- (317,95.5) -- (184.06,95.5) -- cycle ;
\draw   (184.06,95.5) -- (317,95.5) -- (317,127) -- (184.06,127) -- cycle ;

\draw (104.42,103.16) node [anchor=north west][inner sep=0.75pt]   [align=left] {$\pi$-calculus};
\draw (53.89,71.66) node [anchor=north west][inner sep=0.75pt]   [align=left] {Behavioral Equations};
\draw (195.75,102.25) node [anchor=north west][inner sep=0.75pt]   [align=left] {prefix actions, +, $\displaystyle \mid ,\ \nu ,\ !$};
\draw (220.12,71.66) node [anchor=north west][inner sep=0.75pt]   [align=left] {functions, states};

\end{tikzpicture}}
\end{center}
}

\subsection{Semantics of Behavioral Equations}
\label{subsec:semantics}
We translate behavioral equations into $\pi$-calculus process expressions to make precise the \emph{interaction} among concurrent states. Let $\mathcal{B}$ be the set of behavioral equations and $\mathcal{E}$ be the set of $\pi$-calculus process expressions. We define a meta-algorithm $\beta: \mathcal{B} \to \mathcal{E}$ that translates a behavioral equation into a process expression.

\subsubsection{Semantics of Non-recursive Behavioral Equations}
\label{subsec:nonrecursive}
We first consider non-recursive behavioral equations. Let $b \in \mathcal{B}$ be a non-recursive behavioral equation like in \cref{eq:beh}, $p \neq q$. Translating $b$ to $\pi$-calculus is straightforward, shown below using the syntax of the programming language Scala\footnote{Scala keywords are highlighted in \textbf{boldface}.}:
\changem{
\begin{align}
b&\ \textbf{{match}}\ \textbf{{case }} p:=f\{i_1, \ldots, i_n\}.q \textbf{ if } p\neq q\ {{\boldsymbol{\Rightarrow}}}\ \nonumber \\
& p(x).(\nu d)(\nu m)( \label{eq:l1} \\
& \ i_1(m).\overline{d}m.0 \mid \ldots \mid i_n(m).\overline{d}m.0 \mid \label{eq:l2} \\
& \ d(m_1).\ldots d(m_n).(\nu y)\llbracket  y = f(m_1, \ldots, m_n, x) \rrbracket .!\overline{q}y.0) \label{eq:l3}
\end{align}
}
For clarity, we show the translated $\pi$-calculus expression in three lines, labeled as equations \labelcref{eq:l1}, \labelcref{eq:l2}, and \labelcref{eq:l3},\footnote{An alternative presentation is to name the expressions in \cref{eq:l2} and \cref{eq:l3} with process identifiers $L$ and $M$, and showing \cref{eq:l1} as $p(x).(\nu d)(\nu m)(L \mid M)$. These two presentations are identical.} explained below.
\begin{itemize}
    \item \cref{eq:l1} initializes the value of state $p$ and creates new names $d$ and $m$ for processes in \cref{eq:l2} and \cref{eq:l3} to communicate privately;
    \item \cref{eq:l2} receives values from states in the reference set non-blockingly by parallel composition, and forwards received values to \cref{eq:l3} on private channels; and
    \item \cref{eq:l3} computes $f$ and sends the computed value to state $q$ on the RHS. 
    The expression $\llbracket y = f(m_1, \ldots, m_n, x) \rrbracket$ denotes a $\pi$-calculus process that applies $f$ to values $m_1$, $\ldots$, $m_n$, $x$, binding the result to $y$. The encoding details are in~\cite{milner90functions}. The process $!\overline{q}y$ sends $y$ on the channel $\overline{q}$ many times to initialize the state $q$ on the RHS of $b$ and inform other processes of the value of $q$.
\end{itemize}

Non-recursive behavioral equations can be readily composed to model iterative parallel computations of a BSP system. Though a formal proof is beyond the scope of this paper, we illustrate how to model computations in a BSP system by composing behavioral equations in the example below.

\begin{example}
\label{ex:nonrec}
Consider a minimal BSP system with two BSP cores, labelled core 1 and 2, initially in states $p_1$ and $p_2$ with values $5$ and $6$ respectively. Per superstep, each core sends its value to the other core and applies functions $f$ and $g$ respectively to its value and received values. 

For demonstration, we only show computations in two supersteps. Additionally, we assume that cores send and receive messages at the beginning of the first superstep for simplicity.
Computations in this BSP system can be modelled as the following process $S$: 
\begin{align*}
S & \stackrel{def}= (\nu p_5)(\nu p_6)((\nu p_3)(\nu p_4)((\nu p_1)(\nu p_2)(!\overline{p_1}5 \mid !\overline{p_2}6 \mid \llbracket p_1:= f\{p_2\}\\
& .p_3 \rrbracket 
\mid \llbracket p_2:=g\{p_1\}.p_4 \rrbracket) \mid \llbracket p_3:=f\{p_4\}.p_5 \rrbracket \mid \llbracket p_4:=g\{p_3\}.p_6 \rrbracket)).
\end{align*}
The restriction operator in $\nu p_i$ ensures the state $p_i$ is visible only in the corresponding superstep. The process $(!\overline{p_1}5 \mid !\overline{p_2}6)$ initializes states $p_1$ and $p_2$ with values 5 and 6. 
$\llbracket p_1 := f\{p_2\}.p_3 \rrbracket$ denotes the $\pi$-calculus expression translated from the enclosed non-recursive behavioral equation.\footnote{We overload the meaning of the notation $\llbracket \rrbracket$ compared with its usage in \cref{eq:l3}.} 
Each update to a BSP core is captured by a behavioral equation. Equations $\llbracket p_1 := f\{p_2\}.p_3 \rrbracket$ and $\llbracket p_2 := g\{p_1\}.p_4 \rrbracket$  model state updates in the first superstep in cores 1 and 2 respectively. Similarly for the second superstep.
\end{example}

\subsubsection{Semantics of Recursive Behavioral Equations}
\label{subsec:recursive}

Recursive behavioral equations are of the form
$p := f\{i_1, \ldots, i_n\}.p$,
where $p$ is on both the LHS and RHS. This causes problems when {composing} equations in parallel to model a BSP system. For instance, substituting non-recursive behavioral equations in \cref{subsec:nonrecursive} with recursive behavioral equations for $p_1$ and $p_2$ results in process $S'$: 
$$
S' \stackrel{def}= (\nu p_1)(\nu p_2)(!\overline{p_1}5 \mid !\overline{p_2}6 | \llbracket p_1:= f\{p_2\}.p_1 \rrbracket | \llbracket p_2:=g\{p_1\}.p_2 \rrbracket)
$$
\changetxta{
which no longer models computations of the BSP system in \cref{ex:nonrec}. For starters, $S'$ cannot be reduced to an {irreducible form} -- a process expression where no reduction rules defined in \cref{subsec:pi-calculus} can be applied -- after finitely many reductions. $S'$ models computations that run indefinitely, instead of two supersteps.
}

\changetxta{In addition, $S'$ introduces undesirable non-determinism, allowing states $p_1$ and $p_2$ to receive obsolete state values from past supersteps, as opposed to the most recent value:} In \cref{eq:l3}, $!\overline{q}y$ sends the computed value $y$ many times on $\overline{q}$, to initialize the state of  After computing the updated value of $p_1$, the result $y$ is again sent out on the same channel $\overline{p_1}$, in parallel with the previous value $a$ sent out on the same channel.

To address this, we let the system synchronize the BSP states by introducing new actions $yield$ and $resume$. Per superstep, the BSP system sends messages to each state on the channel $\overline{resume}$ with {values} needed by each state. At the end of a superstep, each state sends its name $p$, computed value $y$, and names that it wants to obtain values of, to the system on the channel $\overline{yield}$.\footnote{$yield$ and $resume$ are \emph{polyadic} names -- channels that can send and bind multiple values. Interested readers can refer to~\cite{milner1993polyadic} for a detailed explanation.} The meta-algorithm is updated with the following case expression:
\changem{
\begin{align*}
&\textbf{{case }} p:=f\{i_1, i_2, \ldots, i_n\}.p \ {{\boldsymbol{\Rightarrow}}}\ P, \\
& P \stackrel{def}{=} resume(m_1, \ldots, m_n).p(x). \\
& \ \ (\nu y)\llbracket  y = f(m_1, \ldots, m_n, x) \rrbracket.(\overline{yield}(p, y, i_1, \ldots, i_n).P \mid \overline{p}y.0)
\end{align*}
}
The expression $\llbracket y = f(m_1, \ldots, m_n, x) \rrbracket$ denotes a $\pi$-calculus process expression that encodes the function application of $f$ and binding the result to $y$. The process $\overline{p}y.0$ stores the calculated value to initialize the value of $p$ in the next superstep after receiving $resume$ from the system.

\subsection{Annotations of Behavioral Equations}
Expressing physical properties like data and computation placement allows the system to distinguish between different copies of a value on different machines but sharing the same state name, which is needed for expressing optimizations that exploit data locality.

We distinguish states on the same machine from those on different machines by introducing the \emph{partition} abstraction. Conceptually, a partition is another BSP machine with a set of BSP cores. In this regard, our computational model where states are separated into different partitions, can be described as the slightly generalized \emph{partitioned or hierarchical BSP model}~\cite{hbsp, dbsp, dbsp2, dbsp3}.

Each partition has a unique identifier. We {annotate} each state with its partition identifier $p_i$, expressed as 
$p@p_i$, 
and refer to such equations as \emph{partition-annotated} (or annotated) behavioral equations. Partition-annotated behavioral equations achieve all the requirements in the desiderata in \cref{sec:intro}: We can express non-deterministic concurrent behavior, agent computations, agent interactions, and data and computation placement.
\section{Optimization Example}
\label{sec:optimizations}
In this section, we demonstrate \changetxta{the generality and usability} of behavioral equations by expressing various data-sharing and computation-sharing optimizations 
as transformations over behavioral equations, through a concrete example.

\begin{figure*}
\begin{subfigure}{0.45\columnwidth}
\centering
\begin{tikzcd}[column sep=tiny, row sep=tiny]
& {x_1'}_{op} \arrow[dash, dl, ""] \arrow[dash, d, ""] \arrow[dash, dr, ""] \arrow[dash, drr, ""] & \\
local() \arrow[dash, d, ""] & remote() \arrow[dash, d, ""] & remote() \arrow[dash, d, ""] & remote() \arrow[dash, d, ""] \\
x_1 & x_2 & x_3 & x_4
\end{tikzcd}
    \caption{Visualize $x_1 := op\{x_2, x_3, x_4\}.x_1'$ as a computation tree, where $op(v, m_2, m_3, m_4) = min(v, 1 + min(m_2, m_3, m_4))$. $x_1$ is the reference of state $x_1$ and $x_1'$ is the reference of state $x_1'$. $x_2$ to $x_4$ are state references of neighboring agents.}
    \label{fig:intro-alg}
\end{subfigure}
\hfill
\begin{subfigure}{0.45\columnwidth}
\centering
\begin{tikzcd}[column sep=tiny, row sep=tiny]
& {x_1'}_{op}@p_1 \arrow[dash, dl, ""] \arrow[dash, d, ""] \arrow[dash, dr, ""] \arrow[dash, drr, ""] & \\
local() \arrow[dash, d, ""] & local() \arrow[dash, d, ""] & remote() \arrow[dash, d, ""] & remote() \arrow[dash, d, ""] \\
x_1@p_1 & x_2@p_1 & x_3@p_2 & x_4@p_2
\end{tikzcd}
    \caption{Knowing the graph partition after loading an input graph allows the system to compile local messages away. Here states referenced by $x_1$ and $x_2$ are in graph partition $p_1$, and states referenced by $x_3$ and $x_4$ are in partition $p_2$.}
    \label{fig:intro-nolocal}
\end{subfigure}
\begin{subfigure}{0.45\columnwidth}
\centering
\begin{tikzcd}[column sep=tiny, row sep=tiny]
& {x_1'}_{op}@p_1 \arrow[dash, dl, ""] \arrow[dash, d, ""] \arrow[dash, dr, ""] & {x_2'}_{op}@p_1 \arrow[dash, d, ""] \arrow[dash, dl, ""] \\
local() \arrow[dash, d, ""] & local() \arrow[dash, d, ""] & remote() \arrow[dash, d, ""] \\
x_1@p_1 & x_2@p_1 & {d_1}_{min}@p_2 \arrow[dash, d, ""] \arrow[dash, dl, ""]  \\
& local() \arrow[dash, d, ""] & local() \arrow[dash, d, ""] \\
& x_3@p_2 & x_4@p_2 
\end{tikzcd}
\caption{Merging agents.}
\label{fig:intro-fuse}
\end{subfigure}
\hfill
\begin{subfigure}{0.45\columnwidth}
\centering
\begin{tikzcd}[column sep=tiny, row sep=tiny]
& {x_1'}_{op}@p_1 \arrow[dash, dl, ""] \arrow[dash, d, ""] \arrow[dash, dr, ""] & \\
local() \arrow[dash, d, ""] & local() \arrow[dash, d, ""] & remote() \arrow[dash, d, ""] \\
x_1@p_1 & x_2@p_1 & {d_1}_{min}@p_2 \arrow[dash, d, ""] \arrow[dash, dl, ""]  \\
& local() \arrow[dash, d, ""] & local() \arrow[dash, d, ""] \\
& x_3@p_2 & x_4@p_2 
\end{tikzcd}
    \caption{\changec{Aggregation pushdown}.}
    \label{fig:intro-aggregation}
\end{subfigure}
\begin{subfigure}{\columnwidth}
\centering
\begin{tikzcd}[column sep=tiny, row sep=tiny]
& {x_1'}_{op}@p_1 \arrow[dash, dl, ""] \arrow[dash, d, ""] \arrow[dash, dr, ""] \arrow[dash, drr, ""] & {x_2'}_{op}@p_1 \arrow[dash, dl, ""] \arrow[dash, d, ""] \arrow[dash, dr, ""] &  {x_8'}_{op}@p_1 \arrow[dash, dll, ""] \arrow[dash, dl, ""] \arrow[dash, dr, ""] & {x_9'}_{op}@p_1 \arrow[dash, dl, ""] \arrow[dash, d, ""] \arrow[dash, dr, ""] \\
local() \arrow[dash, d, ""] & local() \arrow[dash, d, ""] & local() \arrow[dash, d, ""] & local() \arrow[dash, d, ""] & local() \arrow[dash, d, ""] & local() \arrow[dash, d, ""] \\
x_1@p_1 & x_2@p_1 & (0, 0)@c & (0, 1)@c & x_8@p_1 & x_9@p_1 
\end{tikzcd}
\begin{tikzcd}[column sep=tiny, row sep=tiny]
{c}_{(1, 2)}@p_1 \arrow[dash, d, ""] \\
remote() \arrow[dash, d, ""] \\
\{[x_3, x_4]\}@p_2 \\
\end{tikzcd}
\caption{Synthesize message data structures and transform computations for remote communication into local computations over the synthesized data structures. \changec{Compared with (b), the $remote$ operators for obtaining values of $x_3$ and $x_4$ when constructing $x_1'$ are changed into $local$, to interact with a locally synthesized message data structure $c$ instead.}}
\label{fig:intro-remote}
\end{subfigure}
\caption{Illustration of how various optimizations transform computation trees.}
\label{fig:intro}
\end{figure*}
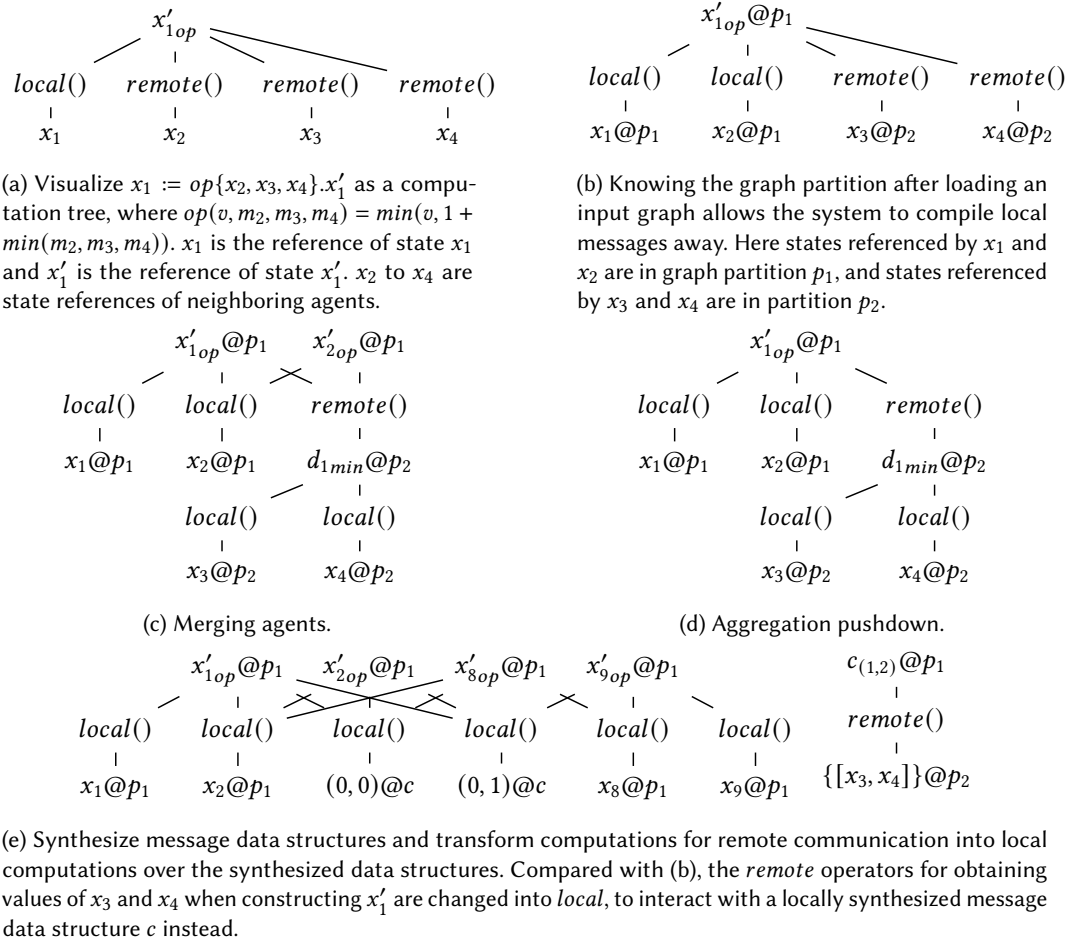

\begin{example}
Let $A_1$ be an agent that aggregates values from agents $A_2$ to $A_4$ and updates its value $v$ to the minimum of $v$ and one plus the minimum of the received values, as described in $op$:
\begin{equation*}
op(v, m_1, m_2, m_3) = min(v, 1 + min(m_1, m_2, m_3)).
\end{equation*}
%
\end{example}
Assume $A_1$ is in state $x_1$, and its neighbors are in states $x_2$, $x_3$, and $x_4$. The behavioral equation for $x_1$ is expressed as:   
\begin{equation}
x_1 := op\{ x_2, x_3, x_4 \}.x_1'.
\label{eq:x1}    
\end{equation}
\Cref{fig:intro-alg} illustrates \cref{eq:x1} as a \emph{computation tree}. The root node is ${x_1'}_{op}$, where $op$ is the user-defined function in \cref{eq:x1}. The leaves consist of names in the reference set together with the state name on the LHS of \cref{eq:x1}. The operators $remote()$ and $local()$ are code generators inserted automatically by the system. Data flows from the bottom up. Nodes at the same height are unordered and can be evaluated independently. Each node evaluates to a state name, allowing such operators to be nested and composed. Below we present various data-sharing and computation-sharing optimizations, and show how they transform \cref{fig:intro-alg}.

\paragraph{Compile Local Messages Away}
\changec{By default, a system assumes that an agent communicates with its neighbors using remote communication primitives, as shown in \cref{fig:intro-alg}. This optimization enables a system to analyze the partition information and rewrite $remote()$ to $local()$ when possible, as illustrated in \cref{fig:intro-nolocal}.}

\paragraph{Merging}
Multiple computation graphs can be merged into one computation graph. Duplicated nodes are removed and their edges are redirected to the unified node, preserving connectivity. This can greatly improve data sharing among nodes in the graph.

\Cref{fig:intro-fuse} shows how to merge computation trees for $x_1$ and $x_2$. The behavioral equation for $x_2$ is defined like follows:
$$
x_2 := op\{x_3, x_4\}.x_2'
$$ 
State $x_2$ aggregates values from $x_3$ and $x_4$ in the same way as $x_1$, before updating its state. 

\paragraph{\changec{Aggregation pushdown}}
This optimization aggregates messages locally at remote machines and sends processed results to an agent, to reduce the amount of intermediate data shuffled in the network and to distribute computations for a better load balance, shown in \cref{fig:intro-aggregation}.\footnote{Applying this optimization requires the computation for processing messages to be commutative and associative, which is satisfied by $min()$ in $op$.}
The operator ${d_1}_{op}@p_2$ creates a state $d_1$ dynamically\footnote{{The term \enquote{dynamic} refers to the fact that the state is created by the system during the optimization phase rather than defined by the user in behavioral equations.}} to store the partial result obtained after processing messages locally in partition $p_2$ and sends the value to $x_1$. 

\paragraph{Synthesize Cross-Partition Message Data Structures}
This optimization minimizes messaging overhead by aggregating communications at the partition level. By analyzing cross-partition edges,\footnote{While this approach uses fixed data structures to manage messages based on a static graph structure, it does not restrict the system to static communication. The system supports dynamic updates, allowing agents to change their references to connected neighbors and request values from new ones. In such cases, it reverts to default agent-to-agent messaging, where messages are exchanged directly between agents.} the optimizer can merge boundary agents for each remote partition and create a fixed data structure, called \enquote{cache}, that serves as a placeholder for the values of boundary agents adjacent to a given partition. These agents are ordered by their ids within the cache to enable efficient offset-based lookups. 

In \cref{fig:intro-remote}, the operator $c_{(1, 2)}@p_1$ creates a cache data structure with a cache reference $c$ in the graph partition $p_1$, as a placeholder for remote messages from partition $p_2$ of size 1$\times$ 2 with message schema $[x_3, x_4]$, shown as $\{[x_3, x_4]\}@p_2$. The first value stored in $c$ is received from $x_3$, and the second is from $x_4$.

\paragraph{Transform Remote Communication into Local Computation}
Finally, a system can transform instructions for remote communication, such as sending or processing messages to and from agents in other partitions, into local computations over locally synthesized message data structures. 

\changec{For instance, in \cref{fig:intro-nolocal},  constructing $x_1'$ needs the value of $x_3$ and $x_4$, which are on different partitions, through the \emph{remote} operators. In \cref{fig:intro-remote}, the \emph{remote} operators are transformed into \emph{local} when constructing $x_1'$ and interact with the locally synthesized message data structure $c$, which is called \emph{local computation}}.

More generally, after synthesizing message placeholders based on the partition structure for the value of cross-partition agents, an optimizer rewrites the computation trees of the boundary agents to interact with the synthesized data structure. A \emph{boundary} agent sends or receives at least one message from agents in other partitions. 

In \cref{fig:intro-remote}, we introduce two additional behavioral equations, 
$$
x_8 := op\{x_2, x_3\}.x_8' \quad x_9 := op\{x_8, x_4\}.x_9'
$$ 
Since $x_3$ and $x_4$ are shared by multiple states and processed differently, we do not push computation to the sender and simply synthesize a cross-partition message data structure $c$ and rewrite the behavior of states $x_1, x_2, x_8, x_9$ to interact with the synthesized message data structure $c$: the behavior of $x_1$ is transformed to look up the value of $c$ at offset $(0, 0)$, shown as $(0, 0)@c$, rather than receiving from $x_3$. The rewrites for $x_8$ and $x_9$ are similar.
\section{{\name}}
\label{sec:implementation}
To demonstrate how behavioral equations can increase data and computation sharing in real stateful parallel systems, we developed OptiFusion, a Scala-based prototype integrated as a library in CloudCity~\cite{tian23generalizing}, a stateful parallel system designed for distributed agent-based simulations.

\Cref{fig:optimizerArch} shows the overall system architecture of {\name}, which has three layers: frontend, optimizer, and backend. In the frontend, users specify agent behaviors using behavioral equations and partition structures. The optimizer then transforms behavioral equations and partitions to improve performance through various optimizations. The optimized partitions are mapped to the agents in CloudCity in the backend for parallel executions.

\begin{figure}
    \centering
    \resizebox{0.7\linewidth}{!}{\tikzset{every picture/.style={line width=0.75pt}} 

\begin{tikzpicture}[x=0.75pt,y=0.75pt,yscale=-1,xscale=1]

\draw   (79.36,22) -- (302,22) -- (302,42) -- (79.36,42) -- cycle ;
\draw   (79.36,42) -- (302,42) -- (302,90) -- (79.36,90) -- cycle ;
\draw   (79.36,90) -- (302,90) -- (302,110) -- (79.36,110) -- cycle ;

\draw (86.01,27) node [anchor=north west][inner sep=0.75pt]  [font=\tiny] [align=left] {\begin{minipage}[lt]{61.11pt}\setlength\topsep{0pt}
\begin{center}
{\tiny Behavioral Equation}
\end{center}

\end{minipage}};
\draw (19, 62) node [anchor=north west][inner sep=0.75pt]  [font=\tiny] [align=left] {\begin{minipage}[lt]{41.99pt}\setlength\topsep{0pt}
\begin{center}
Optimizer
\end{center}

\end{minipage}};
\draw (196.29,27) node [anchor=north west][inner sep=0.75pt]  [font=\tiny] [align=left] {\begin{minipage}[lt]{54.94pt}\setlength\topsep{0pt}
\begin{center}
{Partition Structure}
\end{center}

\end{minipage}};
\draw (19,25) node [anchor=north west][inner sep=0.75pt]  [font=\tiny] [align=left] {\begin{minipage}[lt]{39.44pt}\setlength\topsep{0pt}
\begin{center}
Frontend
\end{center}

\end{minipage}};
\draw (192.23,52) node [anchor=north west][inner sep=0.75pt]  [font=\tiny] [align=left] {\begin{minipage}[lt]{72.35pt}\setlength\topsep{0pt}
\begin{center}
{Synthesize Messages}
\end{center}

\end{minipage}};
\draw (87.49,65) node [anchor=north west][inner sep=0.75pt]  [font=\tiny] [align=left] {\begin{minipage}[lt]{65.1pt}\setlength\topsep{0pt}
\begin{center}
{Rewrite Local Communication}
\end{center}

\end{minipage}};
\draw (160, 96.3) node [anchor=north west][inner sep=0.75pt]  [font=\tiny] [align=left] {\begin{minipage}[lt]{50.59pt}\setlength\topsep{0pt}
\begin{center}
{CloudCity}
\end{center}

\end{minipage}};

\draw (19,96) node [anchor=north west][inner sep=0.75pt]  [font=\tiny] [align=left] {\begin{minipage}[lt]{45.05pt}\setlength\topsep{0pt}
\begin{center}
Backend
\end{center}

\end{minipage}};

\draw (83.17,52) node [anchor=north west][inner sep=0.75pt]  [font=\tiny] [align=left] {\begin{minipage}[lt]{72.1pt}\setlength\topsep{0pt}
\begin{center}
{Partition Transformation}
\end{center}

\end{minipage}};
\draw (193.36,65) node [anchor=north west][inner sep=0.75pt]  [font=\tiny] [align=left] {\begin{minipage}[lt]{66.19pt}\setlength\topsep{0pt}
\begin{center}
{Rewrite Remote Communication}
\end{center}
\end{minipage}};

\end{tikzpicture}}
    \caption{System architecture of {\name}.}
    \label{fig:optimizerArch}
\end{figure}

\subsection{Frontend}
The frontend leverages both non-recursive and recursive behavioral equations.  Users specify the behavior of an agent in one superstep, which is applied repeatedly.\footnote{This is a standard programming model in the vertex-centric paradigm, used by frameworks like Pregel and Giraph.} The iterative computation is expressed by \emph{recursive behavioral equations}, which is modelled by $\syntax{BSP}$ interface. 
\changetxtc{The behavior of an agent is decomposed into fine-grained combinator functions, which synthesize \emph{non-recursive behavioral equations} that describe computations of an agent in a superstep. 
}
Optimizations that transform non-recursive behavioral equations are expressed as type-level operations. 

\paragraph{Behavioral Equation}
A recursive behavioral equation like $p := f\{i, j\}.p$ is represented by a $\syntax{BSP}$ instance, defined below. The name and initial value of the state $p$ are shown on lines 2 and 3 respectively. The reference set is shown on line 4. 

\begin{code}
trait BSP { this: ComputeMethod => 
    val ref: Ref
    var value: Value
    val receiveFrom: Iterable[Ref]
    def run(ms: Set[InMessage]): Unit
}
\end{code}
The $\syntax{run}$ method on line 5 is \emph{generated} by the system and goes through different transformations during the optimization. The agent state value is updated {in-place} in the generated $\syntax{run}$ method. 

The user-defined function $f$ in a behavioral equation is modelled as an instance of $\syntax{ComputeMethod}$ in our programming model. On line 1, the explicit self-type annotation $\syntax{this: ComputeMethod}$ means that an instance of $\syntax{BSP}$ needs to be mixed in~\cite{odersky05scalable} with an stance of $\syntax{ComputeMethod}$. The abstract type members and combinators defined in $\syntax{ComputeMethod}$ are accessible within $\syntax{BSP}$.

\paragraph{User-defined Function}
We model a user-defined function as an instance of $\syntax{ComputeMethod}$, using {combinators} (lines 6--10) to capture the computation structure of the function. Abstract type members (lines 2--4) provide a uniform interface that hides the heterogeneity of type members.

\begin{code}
trait ComputeMethod {
    type Value
    type InMessage
    type OutMessage
    
    def stateToMessage(s: Value): OutMessage
    def partialCompute(ms: Set[InMessage]): Option[InMessage]
    def updateState(s: Value, m: Option[InMessage]): Value
    def deserialize(OutMessage): InMessage
    def run(s: Value, ms: Set[InMessage]): Value = {
        updateState(s, partialCompute(ms))
    }
}
\end{code}

\changec{The $\syntax{stateToMessage}$ combinator (line 6) transforms the value of an agent state into a message before sending it to other agents, to avoid redundant computations by pushing message-processing computations on the sender side. The $\syntax{partialCompute}$ combinator (line 7) enables aggregation pushdown where locally received messages can be aggregated locally into a new message before being sent through the network. The combinator $\syntax{updateState}$ (line 8) computes the updated value based on the current value and aggregated value of received messages declaratively without updating the agent's value in-place, to allow for later optimizations. The $\syntax{deserialize}$ combinator (line 9) allows users to decouple the serialized format of a message from an input value type used for the computation. The $\syntax{run}$ function (line 10) expresses how to compute the new state value, with a default implementation (line 11).
}

\changec{
\paragraph{Example}
We illustrate how to use our new programming model for complex stateful computations via a concrete example, the population dynamics simulation based on Conway's Game of Life.
Each agent represents a grid cell and the world (2D grid) is represented implicitly by a collection of agents. The state of an agent is a Boolean value that denotes whether a cell is alive. The state of an agent evolves in response to the state of its eight adjacent neighbors: 
\begin{itemize}
    \item if an agent has less than 2 or more than 3 alive neighbors, then it dies due to under-population or over-population respectively; 
    \item if an agent has exactly three alive neighbors, then it is alive due to reproduction; 
    \item otherwise, the state of an agent remains unchanged.
\end{itemize}
Per superstep, each agent receives messages from adjacent neighbors that contain their current states and updates its state according to the aforementioned rules. 

\Cref{code:cell-gol} shows how to implement this example using $\syntax{BSP}$ (lines 1--4) and $\syntax{ComputeMethod}$ (lines 6--28). On line 3, $\syntax{FixedCommunication}$ is an annotation in {\name} that indicates the communication pattern remains unchanged across supersteps, which allows for optimizations. 

The combinators in $\syntax{GoLCompute}$ allow the system to easily transform and optimize the program. In this example, each agent needs to count the number of alive neighbors before applying its state update rule. Hence, each agent transforms the value of a neighbor from a Boolean into an integer (1 for true and 0 for false) and sums up the received values. In $\syntax{stateToMessage}$ (lines 26--28), we factor out the transformation from a Boolean into an integer. The $\syntax{partialCompute}$ combinator (lines 11--15) specifies how to partially aggregate the number of alive neighbors. The $\syntax{updateState}$ combinator defines how to apply the update rule. The $\syntax{deserialize}$ combinator is defaulted to the identity function when the $\syntax{InMessage}$ and $\syntax{OutMessage}$ are the same, as is the case here.
}

\begin{figure}
\centering
\begin{code}
class Cell(ref: Ref, neighbors: Seq[Ref]) extends BSP with GoLCompute {
    var value: Boolean = Random.nextBoolean()
    val receiveFrom = FixedCommunication(neighbors) 
}

trait GoLCompute extends ComputeMethod {
    type Value = Boolean
    type InMessage = Int
    type OutMessage = Int

    def partialCompute(m1: Iterable[Int]):Option[Int]= 
        m1 match {
            case Nil => None
            case _ => Some(m1.fold(0)(_+_))
        }

    def updateState(s: Boolean, m: Option[Int]) = 
        m match {
            case None => s
            case Some(n) =>     
                if (n == 3) true 
                else if (n < 2 || n > 3) false  
                else s
        }

    def stateToMessage(s: Boolean): OutMessage = 
        if (s) 1 else 0
}
\end{code}
    \caption{Agent definition for population dynamics example using $\syntax{BSP}$ and $\syntax{ComputeMethod}$ in OptiFusion.}
    \label{code:cell-gol}
\end{figure}

The core of {\name} is to transform a generic agent program like \cref{code:cell-gol} into specialized programs that are efficient to execute by exploiting the partition structure of an input data graph. Next, we explain how to represent the partition structure.

\paragraph{Partition Structure}
In addition to generic agent definitions, users specify an input data graph and its partition structure using $\syntax{Partition}$ when initializing a simulation, shown below.
\begin{code}
trait Partition {
    type NodeId
    type Member

    val id: PartitionId
    val topo: Graph[NodeId]
    val members: Set[Member]
}
\end{code}
A partition has a unique id (line 5) and a set of members (line 7). The graph structure is captured in $\syntax{topo}$ (line 6), where the $\syntax{Graph}$ abstraction contains cross-partition edges. 

\changec{To see how to use the $\syntax{Partition}$ abstraction, we continue with the population dynamics example and show the end-to-end initialization program in \cref{code:part-gol}.\footnote{The initialization program is for a single-machine multi-threaded simulation. In the distributed setting, the initialization program is different.} OptiFusion provides a graph library for generating different graphs, including 2D torus and stochastic block random graph models. On line 1, $\syntax{graph}$ creates an in-memory data graph corresponding to the social network graph of the population dynamics example. After this, users transform each vertex in the data graph into a $\syntax{Cell}$ agent (line 2). The $\syntax{partition}$ function (line 3) is a library function in OptiFusion that separates the data graph into a given number of partitions, where each partition can be easily mapped to a $\syntax{Partition}$ structure (lines 4 -- 11).\footnote{Note that we show the full definition of $\syntax{Partition}$ (lines 4 -- 11) for clarification. In practice, we can also define an implicit function that automatically transforms partitions returned by $\syntax{partition}$ into $\syntax{Partition}$ objects.}
}

Our optimizer exploits the partition structure to generate specialized programs that are efficient to execute. $\syntax{Optimize.default}$ (line 12) transforms a partition through default optimization phases, explained in \cref{subsec:optimizer}. Finally, on line 13, each transformed specialized partition is mapped into an agent in CloudCity through the connector function $\syntax{bspToAgent}$. 

\begin{figure}
\begin{code}
val graph = GraphFactory.torus2D(width, height)
val cells: Map[Int, BSP with ComputeMethod] = graph.adjacencyList.map(i => (i._1, new Cell(i._1, i._2)))
partition(graph, components).zipWithIndex.par.map(i => {
    val part = new Partition {
        type Member = BSP with ComputeMethod
        type NodeId = Ref

        val id = i._2
        val topo = i._1
        val members = i._1.vertices.map(j => cells(j)).toList
    }
    Optimize.default(part)
}).map(i => bspToAgent(part)).seq
\end{code}
\caption{Population dynamics example in OptiFusion: Specifying the partition structure using $\syntax{Partition}$.}
\label{code:part-gol}
\end{figure}

\subsection{Optimizer}
\label{subsec:optimizer}
The optimizer rewrites partitions through a sequence of type-safe transformations, as shown below. 
\begin{code}
trait Optimizer[T <: Partition, V <: Partition] {
    def transform(part: T): V
}
\end{code}
The optimizer is parameterized with two types, $\syntax{T}$ and $\syntax{V}$, both are constrained to extend $\syntax{Partition}$ (as seen from T <: Partition on line 1). The $\syntax{transform}$ function takes a partition of type $\syntax{T}$ as input and returns a partition of type $\syntax{T}$. This design ensures flexibility in handling different partition types while maintaining type safety and facilitating reusable and composable partition transformations.

\begin{figure}[h]
    \centering
    \resizebox{0.7\columnwidth}{!}{
\begin{tikzpicture}[
    node distance=1cm,
    every node/.style={align=center},
    title/.style={font=\footnotesize\bfseries},
    process/.style={rectangle, draw, minimum height=1cm, minimum width=4cm, text width=6cm, align=center},
    action/.style={red, font=\itshape}
]
\node (n1) [title] {BSP with ComputeMethod};
\node (n2) [title, below of=n1] {(BSP with ComputeMethod, (Iterable[Ref], Iterable[Ref]))};
\node (n3) [title, below of=n2] {(BSP with ComputeMethod with Staging, (Iterable[Ref], Iterable[Ref]))};
\node (n4) [title, below of=n3] {(BSP with ComputeMethod with Staging with DoubleBuffer, (Iterable[Ref], Iterable[Ref]))};
\node (n5) [title, below of=n4] {BSP with ComputeMethod};

\draw[->] (n1) -- node[action, right] {Refine Communication} (n2);
\draw[->] (n2) -- node[action, right] {Rewrite Remote Communication\\Synthesize Messages} (n3);
\draw[->] (n3) -- node[action, right] {Rewrite Local Communication} (n4);
\draw[->] (n4) -- node[action, right] {Merging} (n5);
\end{tikzpicture}    
    }
    \caption{Partition transformations in the default optimizer.}
    \label{fig:optimizer}
\end{figure}
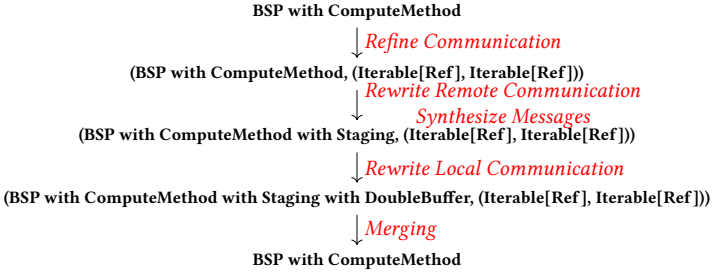

While users are welcome to customize partition transformations, the system provides several built-in transformations, including rewriting local and remote communications and synthesizing message data structures based on partitions. The overall optimization pipeline is shown in \cref{fig:optimizer}.

The first phase in \cref{fig:optimizer} transforms a partition with members of type $\syntax{BSP \ with\ ComputeMethod}$ to a partition with members of type $\syntax{BSP \ with\ ComputeMethod, (Iterable[Ref], Iterable[Ref])}$ after the optimization \emph{refine communication} (shown in red). More specifically, the system filters the connected neighbors for each member in a partition structure to identify (a) local and remote neighbors, and (b) neighbors with static and dynamic communication patterns. Static neighbor references can then be inlined directly into an agent's behavior -- local and remote neighbors generate different instructions -- eliminating repeated lookups. {\name} provides an annotation to let users specify which references in $\syntax{receiveFrom}$ are fixed. After this phase completes, we get a partition where each element contains two parts: First, an adapted version of the original $\syntax{BSP \ with\ ComputeMethod}$ member where its $\syntax{receiveFrom}$ field has been cleaned of static references. Second, a pair called $(localStatic, remoteStatic)$ that captures analyzed neighbor references.

In the next phase, the optimizer synthesizes cross-partition message structures by analyzing inter-partition edges in the graph. The optimizer also generates instructions that read from and write to the generated message structures for references in $remoteStatic$, which otherwise would generate remote communication instructions. The separation of neighbor communication patterns into static and dynamic as well as remote and local, prompted an agent to partially aggregate messages from sharded neighbors. The instructions for partially aggregating values from sharded neighbors are stored as $\syntax{StagedExpr}$ defined in the trait $\syntax{Staging}$. 

Similarly, the optimizer specializes for $localStatic$ by storing computations as staged expressions that are evaluated later. Before doing this, however, the system needs to additionally mix-in the $\syntax{DoubleBuffer}$ trait, which keeps an additional copy of the state value, to ensure that concurrent local accesses should read the correct version of the value, as explained in \cref{sec:pm}. 

Finally, the system can merge a partition, consolidating its members into a single element of a collection type like $\syntax{Vector}$. This can reduce the degree of parallelism and improve data locality, when mapping members of a partition into parallel agents in the distributed backend.

\changec{
    \subsection{Backend}
    The backend of {\name} shares the distributed runtime of CloudCity, which is achieved by integrating {\name} as a library in CloudCity. As seen previously in \cref{code:part-gol}, OptiFusion provides connector functions like $\syntax{bspToAgent}$ that convert a specialized partition into an agent object in CloudCity. \Cref{code:bspToAgent} shows the definition of $\syntax{bspToAgent}$. The attributes $\syntax{id}$ (line 3) and $\syntax{connectedAgentIds}$ (line 4) are built-in attributes for a CloudCity agent. The $\syntax{run}$ method in CloudCity agents~\cite{tian23generalizing} is a co-routine that yields the control to the system at the end of each superstep. The functions $\syntax{toInMessage}$ (line 7) and $\syntax{toOutMessage}$ (line 9) are library functions that transform values in OptiFusion into messages in CloudCity.
}

\begin{figure}[h]
\begin{code}
def bspToAgent(bsp: BSP with ComputeMethod): Actor = 
    new Actor {
        id = bsp.id 
        connectedAgentIds = bsp.receiveFrom

        override def run(): Unit = {
            bsp.run(receivedMessages.map(i => bsp.deserialize(toInMessage[bsp.OutMessage](i))))
            receivedMessages.clear()
            val outMessage = toOutMessage(bsp.stateToMessage(bsp.vaue))
            connectedAgentIds.foreach(n => {
                sendMessage(n, outMessage)
            })}}
\end{code}
\caption{BSP instances in OptiFusion are transformed into agents in CloudCity via functions like $\syntax{bspToAgent}$.}
\label{code:bspToAgent}
\end{figure}
\section{Experiments}
\label{sec:eval}
OptiFusion is a compile-time program specialization framework that exploits a wide range of data-sharing and computation-sharing optimizations. \changetxtm{Our goal is to bridge the performance gap between generic programs, which are easy to develop, and specialized programs, which are efficient to execute through compile-time optimizations.}
We evaluate the effectiveness of {\name}'s optimizations using a benchmark for complex stateful workloads described in \cite{tian23generalizing}. More specifically, we compare the best performance of {\name} with all optimizations enabled, with \changetxtc{four} reference implementations, and show that {\name} achieves:
\begin{itemize}
    \item on par or better performance than the hand-optimized implementations with hard-coded data-sharing and computati\-on-sharing optimizations, and
    \item over 10$\times$ faster than the vertex-centric systems without data or computation sharing optimizations, like Giraph, Flink Gelly, and CloudCity.
\end{itemize} 
Additionally, we provide a detailed breakdown of how each optimization in {\name} contributes to overall performance and report the time spent on the optimizer.

\subsection{Configuration}
Our experiments use 10 servers, each configured with an Intel Xeon Silver 4214 processor, which features 24 cores, 48 hardware threads, a clock frequency of 2.2GHz, and 256GB memory. \changetxtc{Every server is equipped with two 10Gbps Ethernet network interface cards (NIC), configured to use Link Aggregation Control Protocol (LACP). The servers are distributed across multiple racks within the same network. Switches across racks are connected to each other via switches using two 100Gbps Ethernet links.
} The operating system is Debian 12. Software dependencies include OpenJDK 11, Scala 2.12.18, and CloudCity 2.0-SNAPSHOT.

We benchmark different implementations using the agent-based simulation benchmark introduced in~\cite{tian23generalizing}. This benchmark contains a diverse set of workloads: population dynamics, economics, and epidemiology. Details of the workloads can be found in \cite{tian23generalizing}. The population dynamics simulation is the game of life, where each agent is connected to eight adjacent agents. The economics simulation models an evolutionary stock market, where the stock price is seen as the emerging property of traders' buy and sell actions. There is one market agent and the rest are trader agents. The market and traders are connected. Traders do not communicate among themselves. For the epidemics example, two random graph models are considered: the Erd\H{o}s-R\'enyi model (ERM), where each edge is included in the generated graph with probability $p=0.01$, and the stochastic block model (SBM), where vertices are separated into 5 balanced blocks. Two vertices in the same block are connected with probability $p=0.01$. Vertices in different blocks are not connected. 

\begin{figure*}
\begin{subfigure}{0.48\linewidth}
    \centering
    \includegraphics[width=\textwidth]{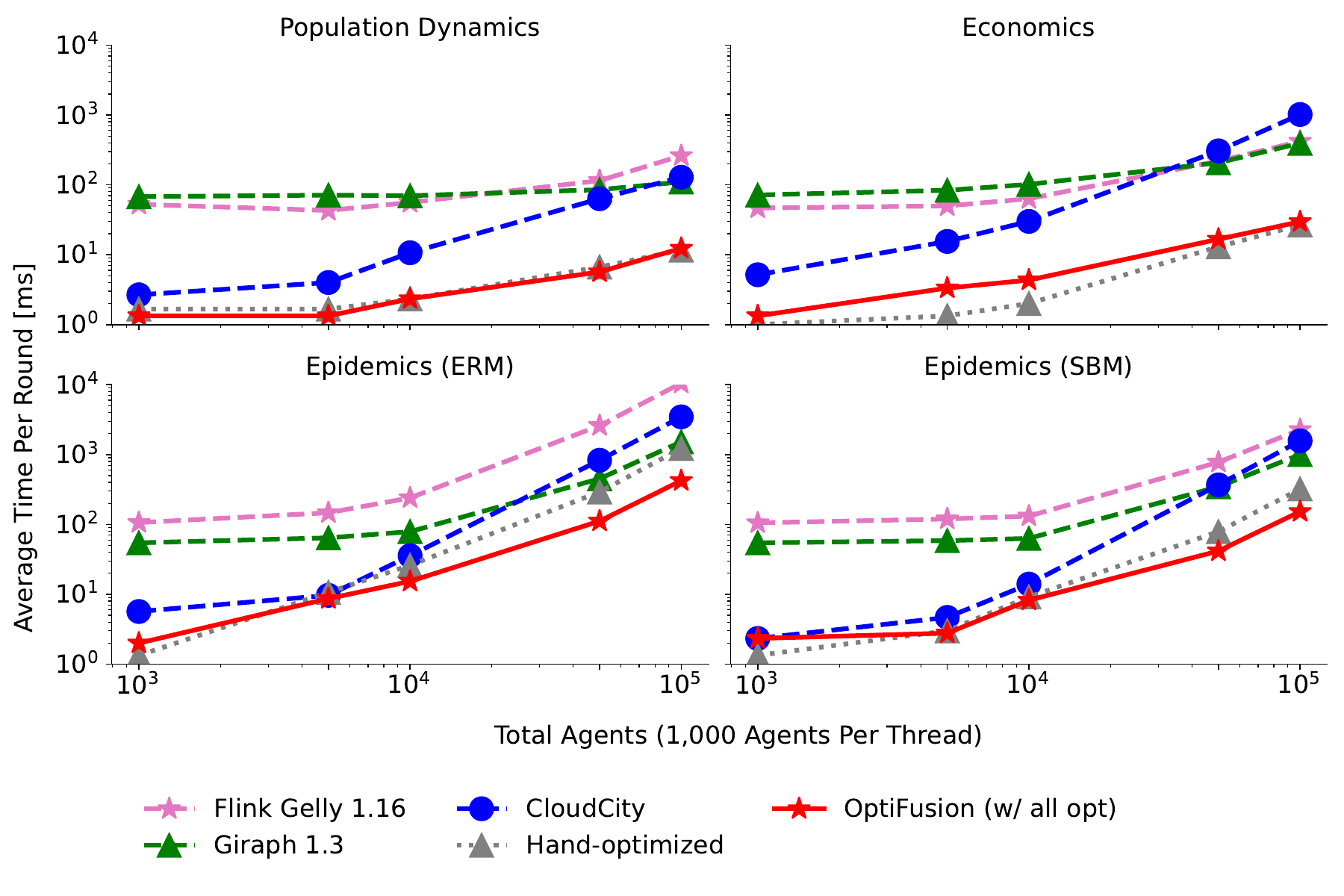}
    \caption{Increase threads (1,000 agents per thread).}
    \label{fig:scaleup}
\end{subfigure}
\hfill
\begin{subfigure}{0.48\linewidth}
    \centering
    \includegraphics[width=\textwidth]{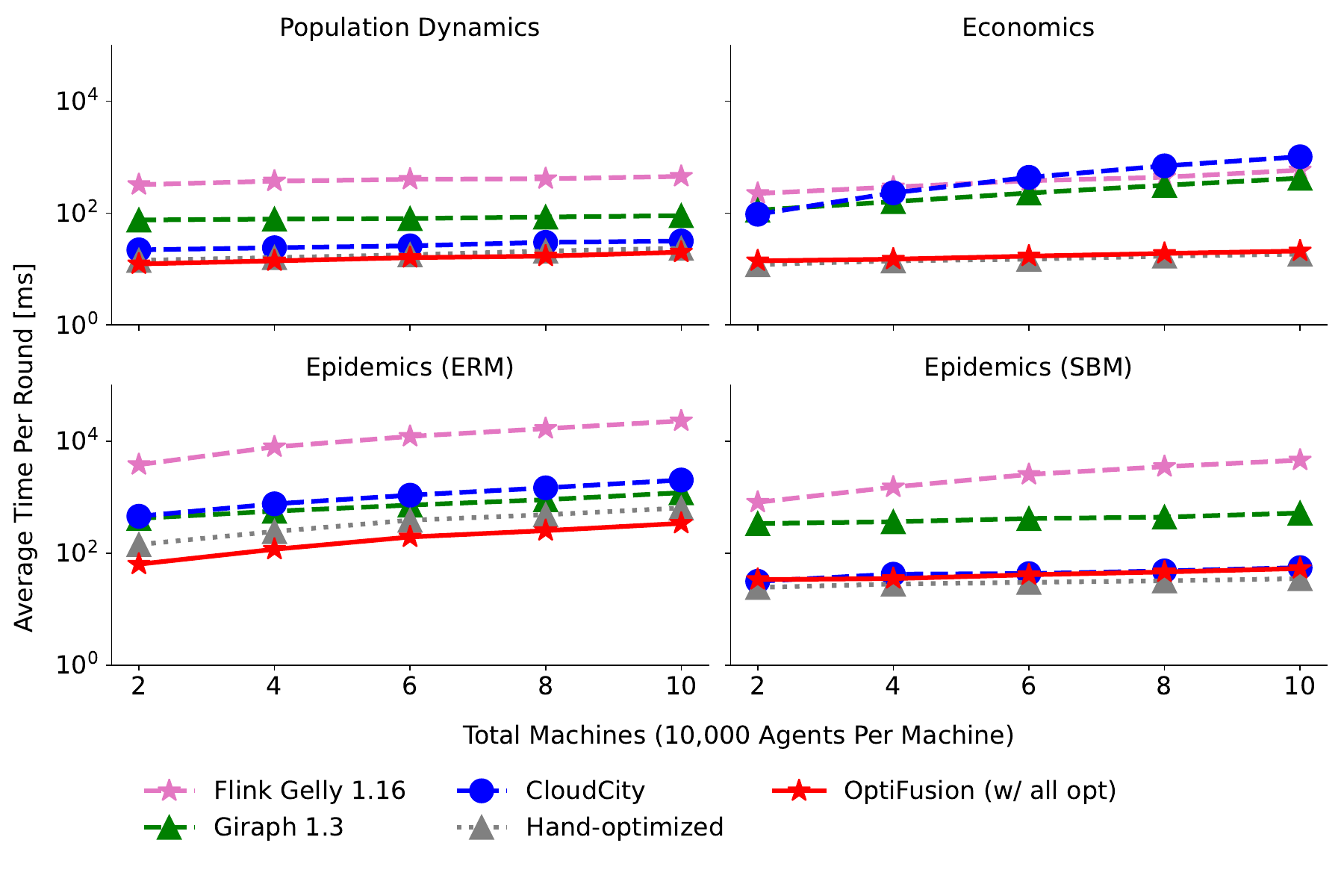}
    \caption{Increase machines (10,000 agents per machine).}
    \label{fig:scaleout}
\end{subfigure}
\caption{\changetxtc{Cross-comparing OptiFusion with other systems} and Hand-optimized when increasing: (a) threads (on one machine), and (b) machines.}
\label{fig:eval}
\end{figure*}

\subsection{Reference Approaches}
\label{subsec:impl-ref}
Recall that in \cref{fig:cross-sys} in \cref{sec:intro}, we presented the experiment results for repeating the scale-up experiments in \cite{tian23generalizing} with the latest software when applicable. We reproduced the result in~\cite{tian23generalizing} and demonstrated that the performance gap between stateless BSP systems and stateful BSP systems remains when executing agent-based simulations. 
\changetxtc{Here we only compare against stateful BSP systems, namely Giraph~\cite{apache-giraph}, Flink Gelly~\cite{flink-pregel}, and CloudCity~\cite{tian23generalizing}. Additionally, we hand-optimize each workload for comparison.}


The system separates the input graph into $K$ balanced components.\footnote{The value of $K$ is determined by the number of threads and the number of machines.} The partitioning strategy affects the number of cross-component messages and is performance-crucial. We evaluate the following strategies: 
\begin{itemize}
    \item Random partitioning: Agents are randomly partitioned into components of a target size.
    \item Hash partitioning: An agent is assigned to a component according to a hash function. We consider the following two self-explanatory hash functions: 
\begin{align*}
&\text{hash}(id) = (id \text{ / } partitionSize).\text{toInt} \\
&\text{hash}(id) = (id \text{ \% } partitionSize).\text{toInt}.
\end{align*}
\item Greedy partitioning: Each component is initialized with a randomly selected agent. We then iteratively add unplaced neighbors of the agents to the component using a breadth-first strategy, continuing until either all neighboring agents are placed or the component reaches its target size.
\end{itemize}

We evaluate each partitioning strategy for the benchmark workloads. Our experiment results\footnote{Omitted here due to space limitations.} showed that the greedy strategy resulted in the best overall performance across all workloads. Since our main goal is to investigate the effectiveness of program transformations that exploit optimizations in {\name}, not how different partitioning strategies affect the performance, we set greedy partition as the default partitioning strategy whenever applicable.

\begin{figure*}
\begin{subfigure}{0.48\linewidth}
    \centering
    \includegraphics[width=\textwidth]{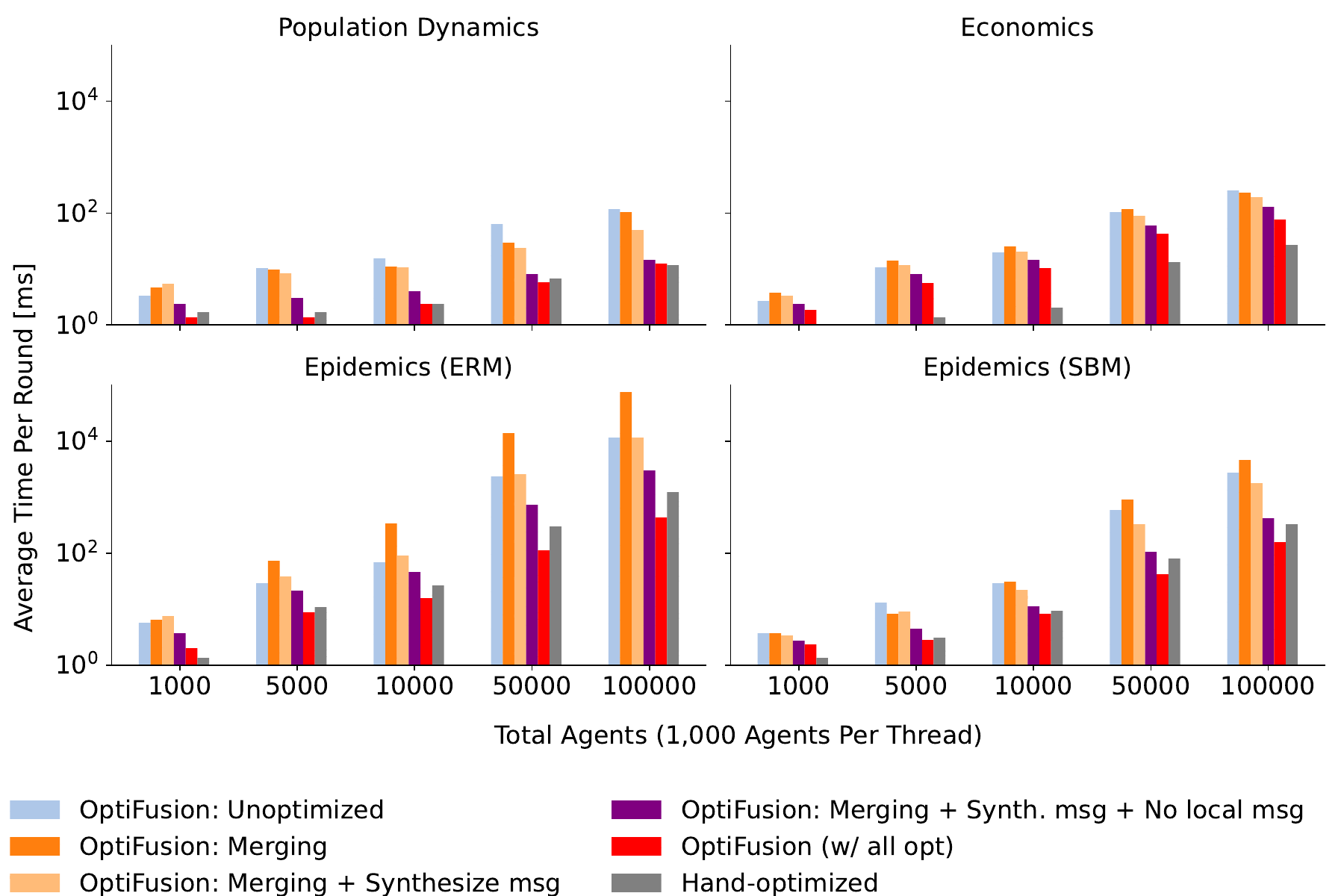}
    \caption{Increase threads.}
    \label{fig:scaleupBreakdown}
\end{subfigure}
\hfill
\begin{subfigure}{0.48\linewidth}
    \centering
    \includegraphics[width=\textwidth]{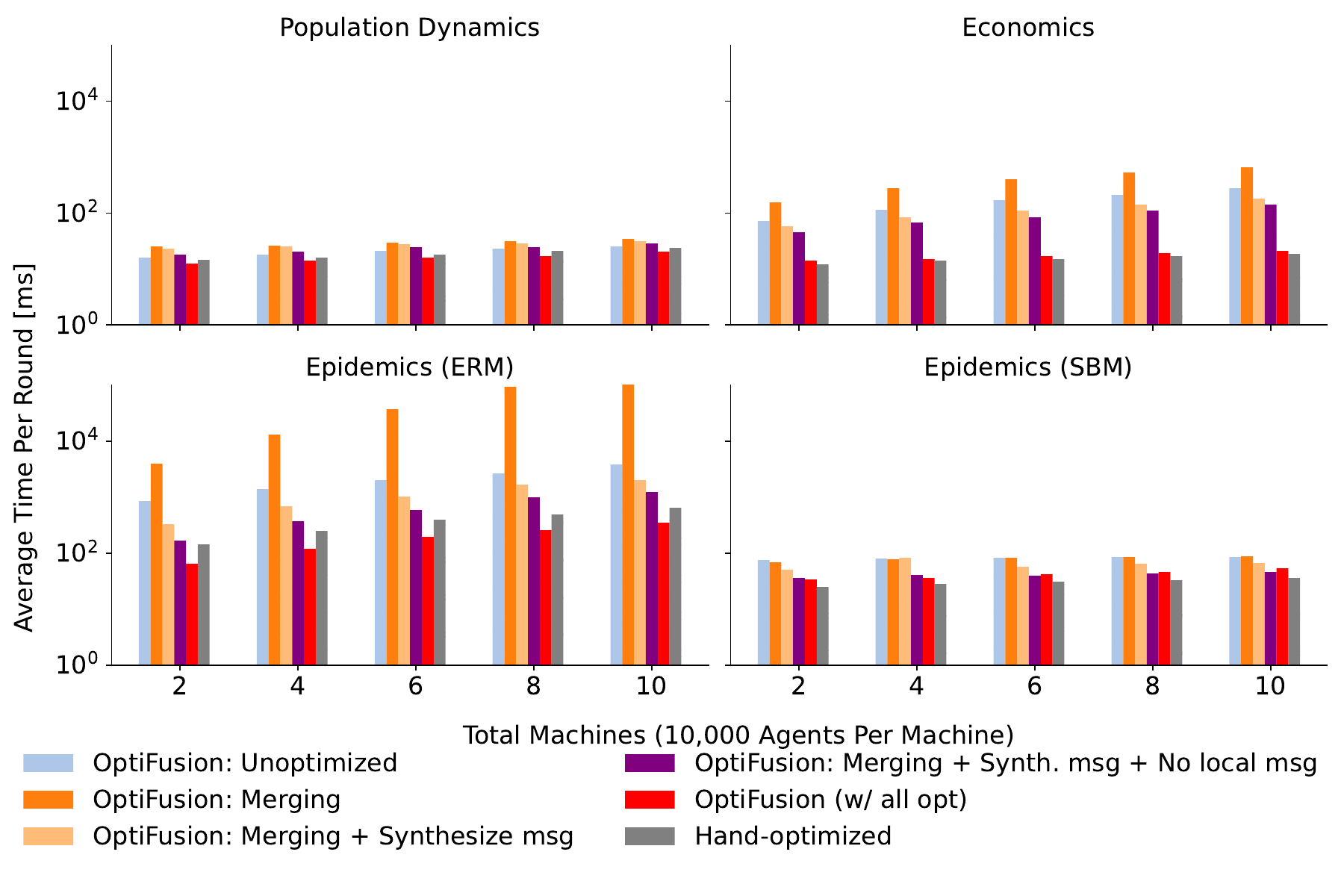}
    \caption{Increase machines.}
    \label{fig:scaleoutBreakdown}
\end{subfigure}
\caption{Analyzing optimization breakdown in {\name} when increasing: (a) threads, and (b) machines.}
\label{fig:eval-breakdown}
\end{figure*}

\subsection{Performance Evaluation}
For the experiments described next, each experiment is repeated three times, and we report the mean of the average time per round for each workload in the benchmark. Each workload is tuned in the same way as documented in~\cite{tian23generalizing}. We run the population dynamics and the economics example for 200 rounds, and the epidemics examples for 50 rounds.\footnote{As explained in \cite{tian23generalizing}, the difference in the number of rounds is to capture \enquote{interesting} computations in the presence of different communication patterns.} 

\changetxtc{To compare the scalability of OptiFusion with other systems}, we fix 1,000 agents per thread and increase the number of threads from 1 to 100 and the number of machines from 1 to 10. Agents are partitioned using the greedy strategy into balanced components that each contain 1,000 agents. Every graph component has a thread. 

\Cref{fig:eval} shows how the average time per round for each workload changes for each system as we increase the number of agents, both by increasing the number of threads and the number of machines. The y-axis is the average time per round measured in milliseconds in the log scale. {\name} achieves comparable or better performance than hand-optimized implementations, both when increasing the number of agents on one machine and when increasing the number of machines. {\name} can be 10$\times$ faster than \changetxtc{CloudCity, Giraph and Flink Gelly. The impact of system design on the benchmark performance of baseline systems has been analyzed thoroughly in \cite{tian23generalizing} and omitted here.}

\Cref{fig:scaleup} shows the scale-up experiments when increasing the number of threads from 1 to 100. Per thread, the number of agents is fixed to 1,000. The x-axis denotes the number of agents in the log scale. As the number of agents increases, the speedup of {\name} and the hand-optimized implementations over \changetxtc{other baseline systems} increases from 2-4$\times$ to 8-15$\times$ across all workloads. A detailed breakdown of how different optimizations affect the performance is shown in \cref{fig:scaleupBreakdown} and \cref{fig:scaleoutBreakdown}, examined in \cref{subsec:optAnalysis}. 

\Cref{fig:scaleout} shows the scale-out experiments when increasing the number of machines up to ten. We fix 10,000 agents per machine and shows the number of machines in the linear scale on x-axis. Compared with other workloads, the economics experiments witness the highest speedup: hand-optimized and OptiFusion are over 50$\times$ faster than other systems. This  is due to \changetxtc{aggregation pushdown}, where an aggregator agent processes values from traders locally before sending the result to the market agent, which drastically reduces the number of messages and balances computations. For population dynamics, the average time per round is nearly the same as the number of machines increases, since computations and remote messages per machine remains unchanged for all systems. 

In some workloads like ERM, we see that {\name} can outperform even hand-optimized implementations by up to 2$\times$. This is due to code specialization. While {\name} generates instructions specialized for each agent, the hand-optimized implementations resemble a generic, interpreted approach with dispatching overhead. 



The interpretation overhead in hand-optimized implementations increases as the number of neighbors for an agent increases. This is evident in \cref{fig:scaleup}. As the number of agents increases from 1,000 to 100,000, the number of neighbors (on average) for an agent in ERM increases from 10 to 1,000, since the edge probability of the ERM model is set to 0.01, and the speedup of {\name} over hand-optimized also increases to 2$\times$. Similarly for the SBM experiments. For the population dynamics workload, each agent has exactly eight neighbors, regardless of the total number of agents. As a result, {\name} and hand-optimized have similar performance even as the number of agents increases. For the economics example, {\name} and the hand-optimized implementations both have \changetxtc{aggregation pushdown}, as we have explained before.

\subsection{Optimization Analysis}
\label{subsec:optAnalysis}
In the unoptimized {\name}, a behavioral equation is transformed directly into a CloudCity agent. Below we analyze the performance impact of different optimizations in {\name}; each transformation is applied after the previous one.

\paragraph{Merging} This transformation merges a collection of agents in the unoptimized {\name} into one agent. After merging, agents within a graph component can obtain the scope information dynamically, such as which neighbors are local (in the same graph component). 
This sets the stage for later optimizations, which exploit the locality information to specialize agent computations.

Merging unoptimized agents can worsen the performance of unoptimized agents, as seen in \cref{fig:eval-breakdown}. There are two main reasons. First, to deliver messages to a neighbor that is fused in a remote component, an agent needs to prepend the component id to a cross-component message, in addition to specifying the id of the receiver agent. This causes more data to be serialized compared with the unoptimized implementation. Second, merging can worsen the computation imbalance among graph components, causing skewed tail latency that slows down the overall average time per round. Luckily, we can mitigate the overhead of merging with optimizations that are enabled, as we will explain shortly.

\paragraph{Synthesize remote message data structures}
After merging, we evaluate the performance impact of synthesizing data structures for messages between graph components. This optimization addresses the messaging inefficiency of merging, where an agent needs to prepend a component identifier in addition to the agent id to every message. In \cref{fig:eval-breakdown}, we see that this optimization mitigates the overhead of merging and achieves similar or slightly better performance than unoptimized across all workloads, due to more efficient message data structures. The performance improvement is more evident in the scale-out experiments in \cref{fig:scaleoutBreakdown} than in the scale-up experiments in \cref{fig:scaleupBreakdown}, since network messages are serialized. For ERM experiments on 10 machines, this optimization improves the performance of merging by over 100$\times$. 

\paragraph{Compile away local messages}
We further consider compiling local messages away, assuming that an agent sends the same value to all neighbors. Instead of materializing and sending these local messages, we let a sender agent generate one message that stores the value of its state. Other agents in the same graph component look up this message locally. Note that we have separated transforming communication instructions into local computation instructions as another optimization. \Cref{fig:eval-breakdown} shows that this technique improves the performance by up to 2$\times$. 

\paragraph{Rewrite communication instructions into local computations}
This optimization inlines the value of neighbor ids into specialized instructions generated for each agent after compiling away local messages, improving the performance by up to 5$\times$, finally allowing {\name} to achieve on par or even better performance than the hand-optimized implementations.

\paragraph{\changec{Aggregation Pushdown}}
The final optimization that we evaluate is \changec{aggregation pushdown}. As explained before, this optimization is only applied to the economics workload. We introduce an aggregator agent in each graph component to combine the actions of traders locally before sending the aggregated value to the market agent. This drastically reduces network messages and results in over 10$\times$ speedup for the economics example.

\subsection{Optimization Overhead}
We have shown that {\name} can match and even surpass the performance of hand-optimized implementations. Here we examine the overhead of our optimizations, namely memory overhead associated with the partition structure and end-to-end time including program transformation time.

\begin{figure}
    \centering
    \begin{subfigure}[b]{0.68\textwidth}
        \centering
        \resizebox{\linewidth}{!}{\input{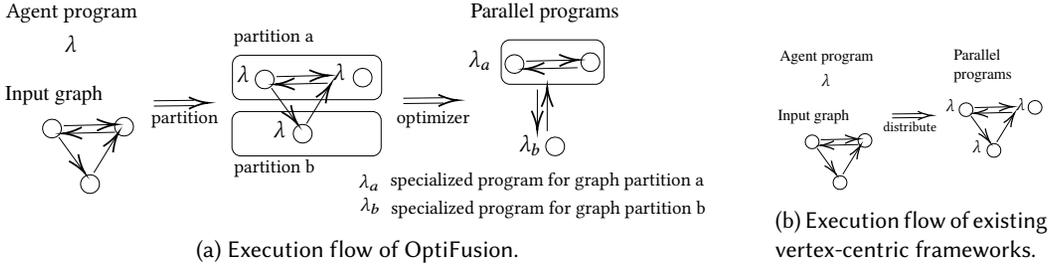}}
        \caption{Execution flow of {\name}.}
        \label{fig:optifusion_runtime}
    \end{subfigure}
    \hfill
    \begin{subfigure}[b]{0.26\textwidth}
        \centering
        \resizebox{\linewidth}{!}{\begin{tikzpicture}[x=0.75pt,y=0.75pt,yscale=-1,xscale=1]

\draw   (46,106.13) .. controls (47.53,103.65) and (50.83,102.84) .. (53.37,104.33) .. controls (55.92,105.82) and (56.74,109.04) .. (55.22,111.53) .. controls (53.69,114.02) and (50.39,114.82) .. (47.85,113.33) .. controls (45.3,111.84) and (44.48,108.62) .. (46,106.13) -- cycle ;
\draw   (65.44,73.75) .. controls (66.96,71.26) and (70.26,70.45) .. (72.81,71.94) .. controls (75.35,73.43) and (76.18,76.66) .. (74.65,79.14) .. controls (73.13,81.63) and (69.83,82.44) .. (67.28,80.95) .. controls (64.74,79.46) and (63.91,76.23) .. (65.44,73.75) -- cycle ;
\draw   (24.29,74.79) .. controls (25.82,72.3) and (29.12,71.5) .. (31.66,72.99) .. controls (34.2,74.48) and (35.03,77.7) .. (33.5,80.19) .. controls (31.98,82.67) and (28.68,83.48) .. (26.13,81.99) .. controls (23.59,80.5) and (22.76,77.28) .. (24.29,74.79) -- cycle ;
\draw    (32.76,82.36) -- (46.55,101.12) ;
\draw [shift={(47.73,102.74)}, rotate = 233.7] [color={rgb, 255:red, 0; green, 0; blue, 0 }  ][line width=0.75]    (10.93,-3.29) .. controls (6.95,-1.4) and (3.31,-0.3) .. (0,0) .. controls (3.31,0.3) and (6.95,1.4) .. (10.93,3.29)   ;
\draw    (65.7,81.65) -- (52.81,102.39) ;
\draw [shift={(66.75,79.95)}, rotate = 121.86] [color={rgb, 255:red, 0; green, 0; blue, 0 }  ][line width=0.75]    (10.93,-3.29) .. controls (6.95,-1.4) and (3.31,-0.3) .. (0,0) .. controls (3.31,0.3) and (6.95,1.4) .. (10.93,3.29)   ;
\draw    (63.44,75.24) -- (35.6,75.88) ;
\draw [shift={(65.44,75.19)}, rotate = 178.67] [color={rgb, 255:red, 0; green, 0; blue, 0 }  ][line width=0.75]    (10.93,-3.29) .. controls (6.95,-1.4) and (3.31,-0.3) .. (0,0) .. controls (3.31,0.3) and (6.95,1.4) .. (10.93,3.29)   ;
\draw    (64.91,79.58) -- (37.07,80.23) ;
\draw [shift={(35.07,80.27)}, rotate = 358.67] [color={rgb, 255:red, 0; green, 0; blue, 0 }  ][line width=0.75]    (10.93,-3.29) .. controls (6.95,-1.4) and (3.31,-0.3) .. (0,0) .. controls (3.31,0.3) and (6.95,1.4) .. (10.93,3.29)   ;
\draw    (91.17,56.04) -- (114.14,55.78)(91.2,59.04) -- (114.17,58.78) ;
\draw [shift={(122.16,57.18)}, rotate = 179.33] [color={rgb, 255:red, 0; green, 0; blue, 0 }  ][line width=0.75]    (10.93,-3.29) .. controls (6.95,-1.4) and (3.31,-0.3) .. (0,0) .. controls (3.31,0.3) and (6.95,1.4) .. (10.93,3.29)   ;
\draw   (165.88,81.38) .. controls (167.41,78.89) and (170.71,78.09) .. (173.25,79.58) .. controls (175.8,81.07) and (176.62,84.29) .. (175.09,86.78) .. controls (173.57,89.26) and (170.27,90.07) .. (167.72,88.58) .. controls (165.18,87.09) and (164.35,83.87) .. (165.88,81.38) -- cycle ;
\draw   (197.91,48.27) .. controls (199.43,45.78) and (202.73,44.97) .. (205.28,46.47) .. controls (207.82,47.96) and (208.65,51.18) .. (207.12,53.67) .. controls (205.6,56.15) and (202.3,56.96) .. (199.75,55.47) .. controls (197.21,53.98) and (196.38,50.75) .. (197.91,48.27) -- cycle ;
\draw   (144.17,50.04) .. controls (145.69,47.55) and (148.99,46.74) .. (151.54,48.23) .. controls (154.08,49.72) and (154.91,52.95) .. (153.38,55.43) .. controls (151.86,57.92) and (148.56,58.73) .. (146.01,57.24) .. controls (143.47,55.75) and (142.64,52.52) .. (144.17,50.04) -- cycle ;
\draw    (152.64,57.61) -- (166.42,76.37) ;
\draw [shift={(167.61,77.98)}, rotate = 233.7] [color={rgb, 255:red, 0; green, 0; blue, 0 }  ][line width=0.75]    (10.93,-3.29) .. controls (6.95,-1.4) and (3.31,-0.3) .. (0,0) .. controls (3.31,0.3) and (6.95,1.4) .. (10.93,3.29)   ;
\draw    (185.57,56.9) -- (172.69,77.63) ;
\draw [shift={(186.63,55.2)}, rotate = 121.86] [color={rgb, 255:red, 0; green, 0; blue, 0 }  ][line width=0.75]    (10.93,-3.29) .. controls (6.95,-1.4) and (3.31,-0.3) .. (0,0) .. controls (3.31,0.3) and (6.95,1.4) .. (10.93,3.29)   ;
\draw    (183.31,50.49) -- (155.48,51.13) ;
\draw [shift={(185.31,50.44)}, rotate = 178.67] [color={rgb, 255:red, 0; green, 0; blue, 0 }  ][line width=0.75]    (10.93,-3.29) .. controls (6.95,-1.4) and (3.31,-0.3) .. (0,0) .. controls (3.31,0.3) and (6.95,1.4) .. (10.93,3.29)   ;
\draw    (184.79,54.83) -- (156.95,55.47) ;
\draw [shift={(154.95,55.52)}, rotate = 358.67] [color={rgb, 255:red, 0; green, 0; blue, 0 }  ][line width=0.75]    (10.93,-3.29) .. controls (6.95,-1.4) and (3.31,-0.3) .. (0,0) .. controls (3.31,0.3) and (6.95,1.4) .. (10.93,3.29)   ;

\draw (35.03,23.52) node [anchor=north west][inner sep=0.75pt]   [align=left] {$\displaystyle \lambda $};
\draw (131.94,43.51) node [anchor=north west][inner sep=0.75pt]   [align=left] {$\displaystyle \lambda $};
\draw (152.68,74.64) node [anchor=north west][inner sep=0.75pt]   [align=left] {$\displaystyle \lambda $};
\draw (186.76,42.07) node [anchor=north west][inner sep=0.75pt]   [align=left] {$\displaystyle \lambda $};
\draw (0.86,51.72) node [anchor=north west][inner sep=0.75pt]  [font=\small] [align=left] {{\fontfamily{ptm}\selectfont Input graph}};
\draw (2.39,5.78) node [anchor=north west][inner sep=0.75pt]  [font=\small] [align=left] {{\fontfamily{ptm}\selectfont Agent program}};
\draw (83.03,60.95) node [anchor=north west][inner sep=0.75pt]  [font=\footnotesize] [align=left] {{\footnotesize distribute}};
\draw (137.96,5) node [anchor=north west][inner sep=0.75pt]  [font=\small] [align=left] {{\fontfamily{ptm}\selectfont Parallel }\\{\fontfamily{ptm}\selectfont programs}};

\end{tikzpicture}}
        \caption{Execution flow of existing vertex-centric frameworks.}
        \label{fig:giraph_runtime}        
    \end{subfigure}
    \caption{Comparison of execution flows between {\name} and traditional frameworks. {\name} partially evaluates an agent program $\lambda$ for a graph partition and generates a special agent program for the partition.}
    \label{fig:runtime_comparison}
\end{figure}

The optimization overhead concerns the preprocessing phase, during which the optimizer of OptiFusion compiles an agent program into a specialized partition program, by partially evaluating the agent program with respect to a graph partition, as shown in \cref{fig:optifusion_runtime}. The generated partition programs are executed in parallel. Other systems, like CloudCity, Giraph, and Flink, do not have such a compilation phase or a partition structure, like shown in \cref{fig:giraph_runtime}. Users input an agent program and a distributed input graph, as shown on the left of \cref{fig:giraph_runtime}, and the framework distributes the agent program to vertices in the input graph and executes agent programs in parallel on all vertices, as shown on the right of \cref{fig:giraph_runtime}.

\changec{\subsubsection{Memory Footprint}
Our optimizer compiles an agent program into a partition program specialized for a graph partition through several stages of rewriting. This can consume more memory than the hand-optimized implementation, due to the auxiliary partition structure as well as the intermediate data structures used in the optimization phase. The memory overhead can be addressed via metaprogramming that modifies the underlying abstract syntax tree of a program, instead of generating a new closure with the transformed program. 

We measure the memory consumption of the optimizer in OptiFusion and compare it with the memory consumption of the hand-optimized implementations prior to the start of simulations, as the number of agents increases from 1,000 to 100,000 on one machine, shown in \cref{fig:memory}. The x-axis shows the number of agents in the log scale and the y-axis shows the memory consumption measured in MB in the log scale. 

\begin{figure}
    \centering
    \includegraphics[width=0.7\linewidth]{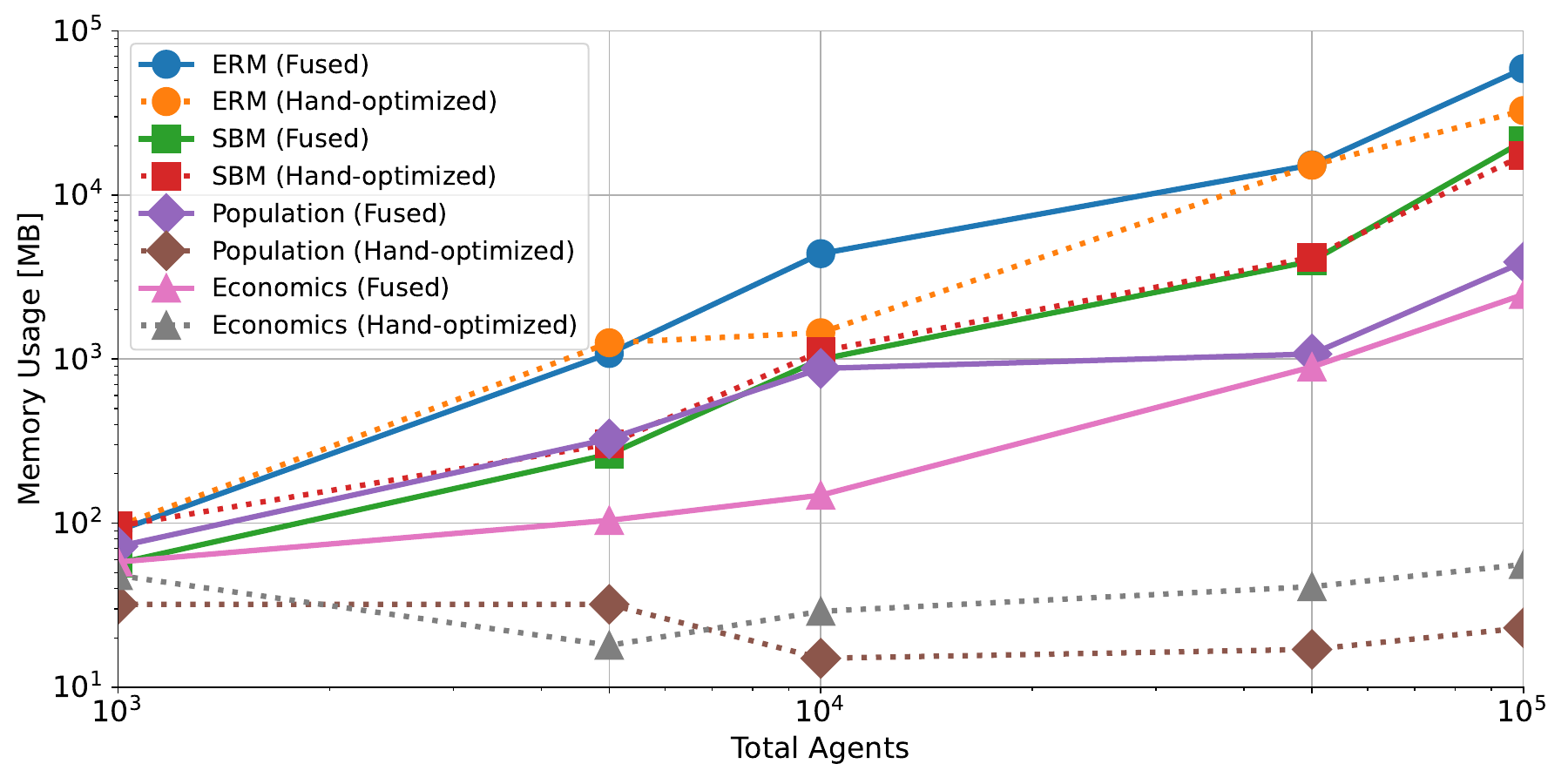}
    \caption{\changetxtc{Memory consumption} of partition structures in OptiFusion and the hand-optimized implementations; other systems do not have partition structures.}
    \label{fig:memory}
\end{figure}

For reference, we also measured the memory consumption of agents -- without auxiliary partition structures -- on other systems. In Flink, there are 50 workers, each configured with 20GB memory. The driver has 50GB memory. As the number of agents increases, the memory consumption of each worker increases from around 8GB to 12GB. The variation across workloads is within 20\%. However, the overall memory consumption is not a simple aggregation. Reducing the number of workers from 50 to 1 does not cause the memory consumption to increase 50$\times$: the memory consumption increases from 10GB to 18GB as the number of agents increases. The memory consumption of agents in Giraph is similar to Giraph, since both systems use Hadoop MapReduce as backend.

\Cref{fig:memory} shows that OptiFusion consumes more memory than the hand-optimized implementation. Though each workload has the same number of agents, their memory consumption varies, closely related to the average number of neighbors per agent. In both OptiFusion and the hand-optimized implementation, the amount of memory consumed by the epidemics examples (ERM, SBM) are greater than other workloads.  

\Cref{fig:memory} also shows that the memory consumption can actually decrease slightly when increasing the number of agents for hand-optimized experiments: from 5,000 to 10,000 for population dynamics, and from 1,000 to 5,000 for economics. This is due to just-in-time (JIT) compilation. To see this, in the population dynamics experiments, the dimension of the grid is 50$\times$100 and 100$\times$100 for 5,000, and 10,000 agents respectively. Hence, the code that describes the behavior of a row of agents is repeated for 50 and 100 times, which can cause the JIT compiler in the Java Virtual Machine to optimize at different levels: the JIT compiler optimizes more aggressively when a code is repeated 100 times than 10 or 50 times. Similarly for the economics experiments. Such effects are more noticeable when the overall memory usage is low.
}

\subsubsection{End-to-End Performance}
\begin{figure}
    \centering
    \includegraphics[width=0.8\linewidth]{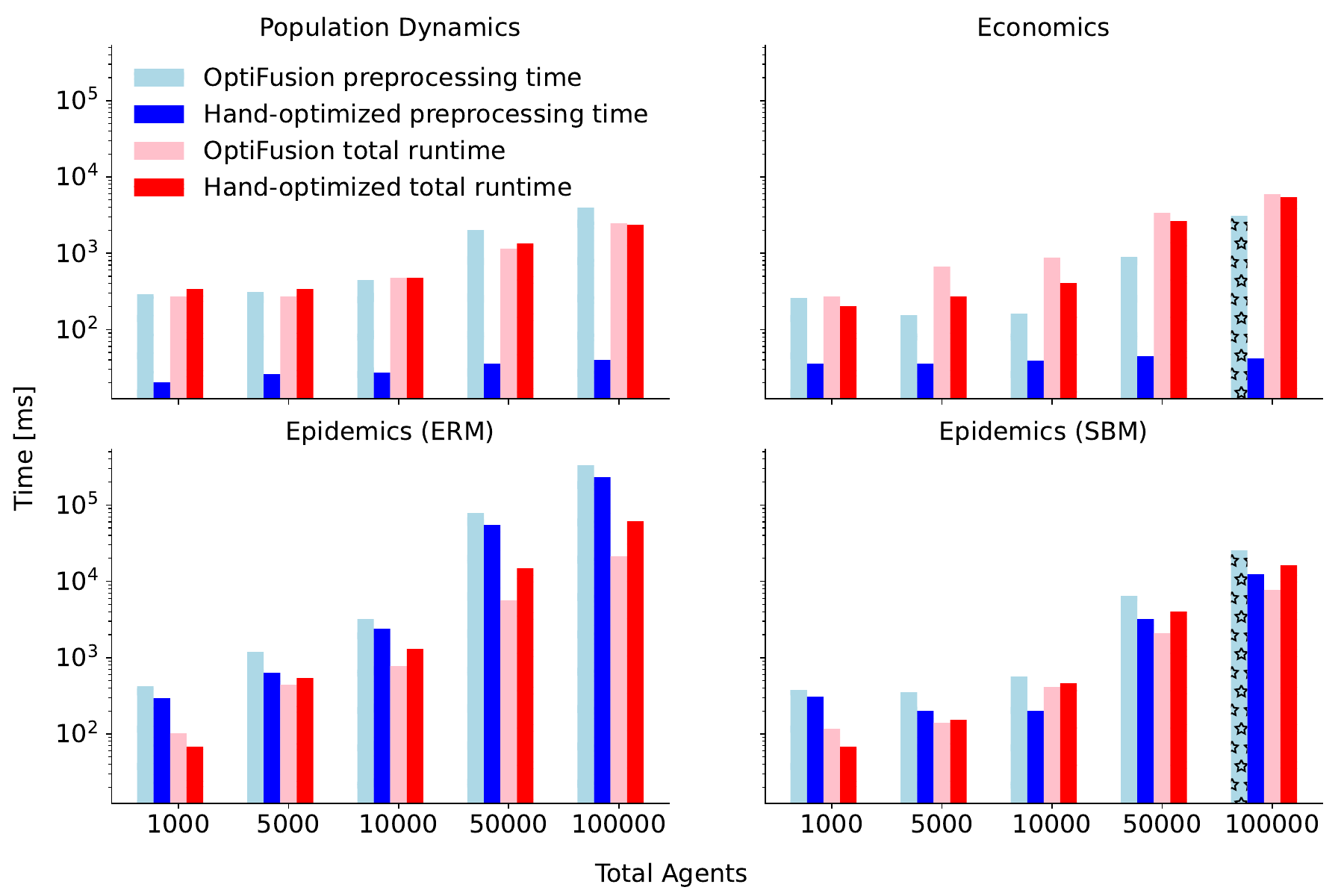}
    \caption{\changetxtc{Comparing end-to-end performance: preprocessing time (fixed, one-time cost) and total runtime (increases when increasing the number of rounds).}}
    \label{fig:endToEnd}
\end{figure}

We also measure the end-to-end time in OptiFusion and compare it with hand-optimized. In addition to runtime, we consider the preprocessing time, including the time to construct the underlying social graph and the compilation time to generate specialized programs, when applicable. While the total runtime depends on the number of rounds, the optimization time spent during preprocessing is one-off and the percentage of optimization time in the overall time decreases as the number of rounds increases.

In \cref{fig:endToEnd}, the preprocessing time in the epidemics examples is longer than their runtime in both OptiFusion and the hand-optimized implementations. This suggests that constructing a social network graph according to random graph models ERM and SBM dynamically is more costly than running the simulations. In these examples, the optimizer overhead is less than 10\%  of the overall preprocessing time; the preprocessing time of OptiFusion is around 10\% longer than that of hand-optimized. The preprocessing time can be reduced by generating the random graph models separately, persisting the generated graph on disk, and reusing these graphs across simulations. For the population dynamics and the economics examples, however, where the social graphs are trivial to generate for hand-optimized implementations, although the percentage of the optimizer overhead remains approximately the same, the overall preprocessing time of hand-optimized is much less than OptiFusion.

We also measured the time breakdown for each optimization phase described in \cref{sec:implementation}. In the first phase, the optimizer refines the $\syntax{receiveFrom}$ values of each agent to identify which communication can be removed, which takes 5\% of the optimizer time. After this, the optimizer synthesizes message data structures for cross-partition messages and rewrites remote communication instructions into local computations that interact with the synthesized message data structures, which accounts for 8\%--10\% of the time. Finally, the optimizer fuses agents in a partition and rewrites local communication into local computations.

The preprocessing time can be reduced by exploiting more efficient algorithms for constructing random graphs. Alternatively, one can also read an input data graph from a file, which is also supported by OptiFusion. The optimizer can also be improved by exploiting low-level parallel primitives rather than the default Scala parallel primitives like $par$ and $seq$, as shown in \cref{code:bspToAgent}. Additionally, the current optimizer uses trait mixins to decompose and rewrite programs, which can generate unnecessary data structures. This issue can be improved using metaprogramming, which rewrites the abstract syntax tree of a program directly without generating the intermediate closure objects, but at the cost of adopting a less user-familiar metaprogramming interface.

\section{Related Work}
The design features listed in the desiderata at the beginning of this paper in \cref{sec:intro} have been explored partially in existing distributed systems. Dedalus~\cite{Alvaro10Dedalus} is a Datalog-like language that emphasizes the model-theoretic perspective of distributed programs, featuring the logical aspects of this desideratum, including state-based dynamics and simple interactions like asynchronous communication. In the Boom project~\cite{Alvaro10Boom}, Overlog~\cite{condie08Overlog} is another Datalog-like language that allows users to specify computation and data placement via {location specifiers}, but not state-based dynamics or interactions. While these Datalog-like languages focus on \emph{declarative} programming that specifies top-down transformations of a parallel collection, our language is grounded in the $\pi$-calculus, a widely used concurrency formalism that can model fine-grained \emph{interactions} between items in a parallel collection bottom-up. Another key difference is that these languages are for asynchronous systems while our language targets BSP systems.

Existing optimizations for BSP systems can be coarsely classified as follows: varying the degree of synchrony and leveraging partitions~\cite{low12distributed, powergraph12, han2015giraph, zhang11priter}, optimizing for different hardware~\cite{shun13ligra, zhang15numa, shun15ligra, Khorasani14Cusha, nurvitadhi14Graphgen}, specializing for target applications~\cite{seastar, sairam14triad, tian23multi-stage, salihoglu14optimizing}, and proposing variants of the vertex-centric model~\cite{TianBCTM13, yan14blogel, simmhan14goffish, xie13grace, zhou14efficient, roy13xstream, yuan14fast, spark, graphx14}. While it is common knowledge that data-sharing and computation-sharing -- an umbrella term that refers to various optimization techniques -- can improve performance, there is no existing work that investigates a principled foundation for expressing and exploiting such optimizations automatically. For example, Spark~\cite{spark} users share computation by explicitly materializing the intermediate result for part of a computation graph in the user program, while Giraph++~\cite{TianBCTM13} users achieve computation-sharing by programming boundary vertices in a graph partition to interact with local vertices and external vertices from other graph partitions differently.

Here we propose a novel approach of using a process calculus to model the concurrent behavior of BSP programs for system optimizations. Process calculi, such as the $\pi$-calculus, are formal frameworks designed to model and analyze the intricate interactions in parallel systems through a rigorous mathematical language. These calculi have been widely utilized across various domains, including programming language design~\cite{PierceT00, occam-pi, turi97Towards, plotkin1981structural, simone1985higher}, concurrent constraint programming~\cite{monjaraz12ghc, victor96Constraints, saraswat1992higher, saraswat89concurrent, smolka94Foundation}, network mobility and security~\cite{abadi1997calculus, Alan03Mcalculus, hennessy2007distributed}. Despite their extensive application in these fields and others~\cite{PatrignaniAC19, wischik05explicit, canetti2001universally, bengtson2011psi, baier2008principles, alexander2008process, FournetG96, UCAM-CL-TR-871}, process calculi have yet to be exploited for system optimization purposes.


\changea{The $\pi$-calculus has inspired numerous language variants. The \emph{Spi Calculus} \cite{abadi1997calculus} and \emph{Applied $\pi$-Calculus} \cite{abadi18applied} extend its capabilities to model cryptographic protocols and verify security properties, while the \emph{Join Calculus} \cite{FournetG00} introduces join patterns for distributed programming. The \emph{Fusion Calculus} \cite{ParrowV98} generalizes $\pi$-calculus with symmetric communication, and the \emph{Ambient Calculus} \cite{CardelliG00} focuses on mobile computation through bounded spaces. \emph{Psi-Calculi} \cite{bengtson2011psi} provides a highly customizable framework for diverse applications, and the \emph{Session $\pi$-Calculus} \cite{Honda93} formalizes structured communication via sessions. Behavioral equations are designed to model and transform distributed computations for system optimizations. To the best of our knowledge, our work is the first to demonstrate that process calculi are a valuable and practical optimization tool for system designers. We showed that {behavioral equations} can be used as a uniform framework to express a wide range of practical data and computation sharing techniques.}
\section{Conclusions and Future Work}
In this work, we have introduced a new language called \emph{behavioral equations} based on the $\pi$-calculus and showed how various data-sharing and computation-sharing optimizations can be expressed as transformations of behavioral equations. We have also built the {\name} system based on behavioral equations and demonstrated the effectiveness of optimizations empirically through thorough experiments.

The optimizations we have exploited only begin to tap into the potential of behavioral equations and process calculus. For example, the full flexibility of using $\pi$-calculus, in particular the replication operator, allows the system to distinguish values that are available exactly once from those that are always available. Though not leveraged by the optimizations here, such features are important for expressing privacy and security constraints. The theoretical aspect of behavioral equations, such as expressiveness, can also be further investigated.

We believe that using process calculus, which models fine-grained interactions between processes, provides a clean, generic, and flexible foundation for parallel programming. Our formalism paves the way for future work like analyzing the theoretical properties of stateful systems, such as the correctness condition, the equivalence relation, the complexity of states, and the relationship with other stateless or asynchronous distributed systems.

\begin{acks}
This work was supported by a Postdoc Grant at the University of Zurich, project number FK-24-020.
\end{acks}

\bibliographystyle{ACM-Reference-Format}
\bibliography{ref.bib}

\end{document}